\documentclass[manuscript]{emulateapj}
\usepackage{graphicx}
\usepackage{amssymb}

\newcommand{\ha}{H$\alpha$}
\newcommand{\hb}{H$\beta$}
\newcommand{\hg}{H$\gamma$}
\newcommand{\mic}{$\mu$m}

\begin{document} 

\title{ The WFC3 Infrared Spectroscopic Parallel (WISP) Survey \footnote{Based on observations made with the NASA/ESA Hubble Space Telescope,
which is operated by the Association of Universities for Research in
Astronomy, Inc., under NASA contract NAS 5-26555. These observations are
associated with program 11696} }

\author{Atek, H.\altaffilmark{1}, 
Malkan, M.\altaffilmark{2},
McCarthy, P.\altaffilmark{3}, 
Teplitz, H.I.\altaffilmark{4}, 
Scarlata, C. \altaffilmark{1},
Siana, B. \altaffilmark{5},
Henry, A. \altaffilmark{6},
Colbert, J.W. \altaffilmark{1},
Ross, N.R.\altaffilmark{2},
Bridge, C.\altaffilmark{5},
Bunker, A.J.\altaffilmark{7},
Dressler, A. \altaffilmark{3},
Fosbury, R.A.E. \altaffilmark{8},
Martin, C. \altaffilmark{6},
Shim, H. \altaffilmark{1}
}
\altaffiltext{1}{Spitzer Science Center, Caltech, Pasadena, CA 91125}
\altaffiltext{2}{Dep't. of Physics and Astronomy, Univ. of Calif. Los Angeles}
\altaffiltext{3}{Observatories of the Carnegie Institution for Science, Pasadena, CA 91101} 
\altaffiltext{4}{Infrared Processing and Analysis Center, Caltech, Pasadena, CA 91125}
\altaffiltext{5}{Department of Astronomy, Caltech, Pasadena, CA 91125}
\altaffiltext{6}{Dep't. of Physics, Univ. of Calif. Santa Barbara, CA 93106}
\altaffiltext{7}{Department of Physics, University of Oxford, Denys Wilkinson Building, Keble Road, OX13RH, U.K.}
\altaffiltext{8}{Space Telescope - European Coordinating Facility, Garching bei M\"unchen, Germany}

\begin{abstract}

  We present the WFC3 Infrared Spectroscopic Parallel (WISP) Survey. WISP is obtaining slitless, near-infrared grism spectroscopy of $\sim 90$\ independent, high-latitude fields by observing in the pure parallel mode with Wide Field Camera-3 on the {\it Hubble Space Telescope} for a total of $\sim 250$ orbits. Spectra are obtained with the G102 ($\lambda=0.8-1.17~\mu$m, R$\sim210$) and G141 grisms ($\lambda=1.11-1.67~\mu$m, R $\sim$130), together with direct imaging in the J- and H-bands (F110W and F140W, respectively). 
In the present paper, we present the first results from 19 WISP fields, covering approximately 63 arcmin$^{2}$. For typical exposure times ($\sim6400$~s in G102 and $\sim2700$~s in G141), we reach $5\sigma$\ detection limits for emission lines of $f\sim5\times 10^{-17}$~ergs s$^{-1}$~cm$^{-2}$\ for compact objects. Typical direct imaging 5$\sigma$-limits are 26.8 and 25.0 magnitudes (AB) in F110W and F140W, respectively. Restricting ourselves to the lines measured with highest confidence,
we present a list of 328 emission lines, in 229 objects, in a redshift range $0.3 < z < 3$. The single-line emitters are likely to be a mix of H$\alpha$ and [OIII]5007,4959 \AA, with H$\alpha$ predominating. The overall surface density of high-confidence emission-line objects in our sample is approximately 4 per arcmin$^{2}$.
These first fields show high equivalent width sources, AGN, and post starburst galaxies. The median observed star formation rate of our \ha\ selected sample is 4 M$_{\odot}$ year$^{-1}$. At intermediate redshifts, we detect emission lines in galaxies as faint as $H\textunderscore140 \sim 25$, or $M_R < -19$, and are sensitive to star formation rates down to less than 1 M$_{\odot}$ year$^{-1}$. The slitless grisms on WFC3 provide a unique opportunity to study the spectral properties of galaxies much fainter than $L^*$ at the peak of the galaxy assembly epoch.

\end{abstract}
\maketitle

\section{Introduction}\label{sec:intro}

While there is considerable uncertainty regarding the detailed star
formation history of galaxies of various masses and types, it has
become clear that the span of cosmic time $0.5 \le z \le 2.5$
encompasses most of the star formation in the history of the Universe \citep[e.g.][]{hopkins04,daddi07}.
This broad epoch is also when many galaxies were ``assembled,'' when
major mergers and the accretion of satellite galaxies and gas combined
to produce the variety of morphologies and structures of modern
galaxies.  There is mounting evidence for an evolution of star-forming galaxies during this
time period, from
high-mass galaxies dominating at $z \sim 3$, to low-mass galaxies at
the present.  Deep, wide IR surveys are required to cover much of the
peak epoch of star formation, because of the redshifting of the key
diagnostic spectral features of stellar evolution, and to
minimize the effect of obscuration by dust.

The difficulty of obtaining observed-frame near-infrared spectra
causes much of our empirical perspective on galaxy evolution to be
biased by the way in which spectroscopic samples are selected.  Color
selection reliably selects common objects, but misses unexpected
ones; magnitude-limited samples necessarily favor bright-continuum objects.  Ground-based searches for emission lines from faint
high-redshift galaxies are severely impacted by the bright NIR
airglow.  Excluding OH terrestrial airglow frequencies restricts
redshift coverage and compromises redshift identification.
Space-based spectroscopic surveys offer significant advantages for
unbiased target selection. The grisms on the Near Infrared Camera and Multi Object Spectrometer (NICMOS), the Space Telescope Imager and Spectrograph (STIS) and the Advanced Camera for Survey (ACS) have
demonstrated the power of spectroscopy from above the atmosphere,
where the low background emission permits slitless observing \citep{gardner98, pmc99, teplitz03a, pirzkal04,drozdovsky05}. Furthermore, the terrestrial wavelength gaps can only be filled from
space, as first demonstrated by the pure-parallels program
with the NICMOS G141 grism \citep{pmc99, yan99, shim09}.

Much as ground-based objective prism surveys (e.g. Gallego et al. 1999) were used to define flux-limited samples of H$\alpha$ emitters at $z
\sim 0$, the NICMOS grism observations produced complete samples of
emission-line objects at intermediate redshifts. Yan et al. (1999) and Hopkins et al. (2000) used observations with the NICMOS H-band (G141)
grism to derive H$\alpha$ luminosity functions, and to show that the
luminosity density, and hence star formation rate density, has evolved
by a factor of $\sim 10$ from $z \sim 1.4$ to the present. Shim et al. (2009) used a larger sample to derive improved H$\alpha$ luminosity functions in two redshift bins and confirmed the trends
seen in the analysis by Yan et al. and Hopkins et al. The complete reductions/extractions of grism data for the principal modes of NICMOS were released in the Hubble Legacy Archive \citep[HLA,][]{freudling08}. Nearly all of the published results from the NICMOS grisms were based on data with
the G141 H-band disperser. The poor sentivity of the NICMOS detector
at short wavelengths limited the utility of the J\textunderscore110-band grism.

The grisms on ACS have been used to discover faint Ly$\alpha$ emission sources and faint Lyman continuum break sources at $z \sim 3 - 4$\ \citep{malhotra05,pirzkal07} as well as to derive spectrophotometric redshifts for a large sample of
faint galaxies \citep[e.g.][]{straughn08,straughn09}. Extremely deep
ACS grism observations have shown the power of low resolution
spectroscopy above the atmosphere to detect spectral breaks indicative
of passively evolving systems \citep[e.g.][]{daddi05}\ or active star
forming galaxies \citep{pirzkal04,rhoads05,malhotra05,pasquali06}. The full release of the reduced ACS grism data in the HLA is underway \citep{kuemmel09b}. The
STIS slitless modes work at higher resolution than the NICMOS and ACS
grisms, but the relatively low sensitivity of STIS has limited its
ability to sample populations beyond the reach of ground-base
spectroscopic surveys \citep{teplitz03b}.

Wide Field Camera Three (WFC3) on Hubble has a larger field of view,
significantly better detectors and superior sampling compared to
camera 3 on NICMOS. The net improvement in survey speed is
approximately a factor of 20.  This enhanced survey speed does not
fully capture the gains provided by WFC3. The quality of the spectra,
in terms of resolution, a well-focused point spread function, and uniform
sensitivity are also greatly superior to those offered by NICMOS.

In HST's Cycle 17, we began a new observing program of WFC3 pure
parallels, the WFC3 Infrared Spectroscopic Parallel (WISP) survey.  In
$\sim 250$\ orbits, the program will obtain near-infrared slitless
spectroscopy of $\sim 90$\ uncorrelated high latitude fields
(PI=Malkan, GO-11696). This will produce a large sample of emission-line galaxies to measure the star formation rate continuously from at $z \sim 0.5$ to $z \sim 2.5$, over which ground-based searches are severely limited. The study of dust and metallicity in a large, unbiased sample of galaxies and the evolution of the star formation density are among the main goals of our project. In this paper, we present the analysis of the
first 19 fields, from which we have selected high-confidence emission
lines.  In Sections \ref{sec:obs}\ and \ref{sec:datared}, we describe
the survey and data reduction.  We discuss the results in Section
\ref{sec:results}, and finally offer some thoughts on the future
possibilities in Section \ref{sec:discuss}.  Throughout, we assume a
$\Lambda$-dominated flat universe, with $H_0=71$\ km s$^{-1}$\
Mpc$^{-1}$, $\Omega_{\Lambda}=0.73,$\ and $\Omega_{m}=0.27$. All magnitudes are in AB system.

\section{Observations}\label{sec:obs}

The infrared channel of the WFC3 utilizes a 1K$\times$ 1K HgCdTe
detector array.  At a plate scale of 0.13 arcsec px$^{-1}$, the total
field of view covered is 123$''\times$136$''$.  There are
two IR grisms on WFC3: G102 covering the $0.8 - 1.1$ \mic\ range with
a dispersion of 0.0024 \mic/pixel and G141 covering the $1.07 - 1.7$
\mic\ range with a dispersion of 0.0046 \mic/pixel. These correspond
to average resolving powers for point sources of $R = 210$ for the G102 and $R
= 130$ for the G141. For an unresolved line, the FWHM will be about 2 pixels. The details of the grisms and their
characteristics can be found in \cite{kimble08}\ and on the WFC3
website\footnote{http://www.stsci.edu/hst/wfc3/}.


Each target field was observed briefly in direct imaging mode, using
the broadband filter that matches the grism spectral coverage most
closely: F110W for the G102, and F140W for the G141.
As detailed in the next section, all of our observations were carried
out in the ``pure-parallel'' mode with the Cosmic Origins Spectrograph
(COS) or STIS operating as the prime instrument. 

\begin{deluxetable*}{l c c c c c c c c c c c}
\tabletypesize{\small}
\tablecolumns{9}
\tablewidth{0pt}
\tablecaption{Schedule of Observations to date for the WISP Survey \label{tab:schedule}}
\tablehead{
\colhead{Field} & \colhead{   Date} & \colhead{   RA} & \colhead{DEC} & \colhead{Number of} & \colhead{F110W} & \colhead{G102} & \colhead{F140W} & \colhead{G141}& \colhead{Primary} & \colhead{Primary} & \colhead{Primary}  \\
\colhead{ } & \colhead{ } & \colhead{[HMS]} & \colhead{[DMS]} & \colhead{Orbits} & \colhead{[Sec]} & \colhead{[Sec]} & \colhead{[Sec]} & \colhead{[Sec]} & \colhead{Proposal ID} & \colhead{Visit No.} & \colhead{Instrument}
}
\startdata
1  & 2009/11/24 & 01 06 35.29 & $+$15 08 53.8 & 4 & 884 & 4815 & 506 & 2609     & 11720 & 01 & COS \\
5  & 2009/12/20 & 14 27 06.64 & $+$57 51 36.2 & 5 & 1034 & 5515 & 1034 & 5515   & 11720 & 13 & COS \\   
6  & 2009/12/24 & 01 50 17.18 & $+$13 04 12.8 & 4 & 609 & 3609 & 862 & 5015     & 11727 & 09 & COS \\         
7  & 2009/12/25 & 14 27 05.91 & $+$57 53 33.7 & 7 & 834 & 6318 & 1112 & 6224    & 11720 & 14 & COS \\          
9  & 2009/12/29 & 12 29 44.31 & $+$07 48 23.5 & 4 & 759 & 4612 & 684 & 3712     & 11703 & 01 & STIS \\
10 & 2010/01/02 & 09 25 07.84 & $+$48 57 03.0 & 3 & 631 & 3909 & 406 & 2209     & 11742 & 06 & COS  \\ 
11 & 2010/01/03 & 11 02 17.38 & $+$10 54 25.4 & 3 & 556 & 3709 & 456 & 2006     & 11742 & 07 & COS  \\        
12 & 2010/01/04 & 12 09 25.25 & $+$45 43 19.8 & 5 & 1312 & 8221 & 606 & 3009    & 11741 & 04 & COS  \\  
13 & 2010/01/09 & 01 06 38.77 & $+$15 08 26.2 & 3 & 556 & 3009 & 506 & 2409     & 11720 & 03 & COS  \\
14 & 2010/01/09 & 02 34 56.80 & $-$04 06 54.5 & 4 & 834 & 6215 & 481 & 2809     & 11741 & 03 & COS  \\        
15 & 2010/02/12 & 14 09 42.47 & $+$26 21 56.0 & 5 & 1612 & 8321 & 531 & 2609    & 11741 & 26 & COS  \\         
16 & 2010/02/16 & 02 34 54.72 & $-$04 06 42.5 & 5 & 1087 & 6921 & 584 & 2509    & 11741 & 01 & COS  \\  
17 & 2010/02/18 & 02 13 38.11 & $+$12 54 59.3 & 3 & 534 & 3409 & 559 & 3409     & 11727 & 11 & COS  \\
18 & 2010/02/19 & 12 29 17.25 & $+$10 44 00.6 & 5 & 534 & 3512 & 762 & 3824     & 11561 & 04 & COS  \\        
19 & 2010/02/20 & 02 34 54.29 & $-$04 06 30.5 & 5 & 1187 & 8721 & 484 & 2809    & 11741 & 02 & COS  \\         
20 & 2010/02/21 & 14 09 41.15 & $+$26 22 15.1 & 7 & 1815 & 8430 & 559 & 2812    & 11741 & 25 & COS  \\  
21 & 2010/03/05 & 09 27 55.77 & $+$60 27 05.3 & 1 & 0 & 0 & 353 & 2006          & 11728 & 10 & COS  \\ 
22 & 2010/03/05 & 08 52 44.99 & $+$03 09 09.6 & 1 & 0 & 0 & 253 & 1806          & 11728 & 13 & COS  \\ 
23 & 2010/03/06 & 09 43 16.12 & $+$05 27 37.1 & 3 & 0 & 0 & 681 & 4115          & 11598 & 42 & COS  \\          
\enddata
\tablecomments{A summary of the observations performed under the HST
  program GO/PAR 11696 for the WISP Survey to date. Four additional
  fields were deferred
  to a later paper, because they require additional reductions. Note
  that some targets have more 
than 5 orbits of observations.  These are not necessarily all full
orbits, however, 
which is why, for example, a three-orbit observation may have more
than half the total integration 
time of a six-orbit observation, and targets with the same number of orbits have varying total integration times.
}
\end{deluxetable*}


\subsection{Pure Parallel Observation Scheduling} 

Pure-parallel observing with HST works as follows: Observing programs
approved on COS and STIS constitute a parallel-observing `opportunity'
(see Table~\ref{tab:schedule}).  Observations with these two
instruments, by necessity, require stable, long integrations, which
means that the entire focal plane of HST is held at a single pointing
and fixed orientation angle for long periods of time.  Thus it is simply a
matter of turning on one of the other instruments in the focal plane
to obtain quality data in a parallel field, offset by a 5.5\arcmin\
and 4.75\arcmin\ from the COS and STIS primary target,
respectively. This motivated the strategy of a number of programs
using WFC3 in parallel. Two other HST Cycle 17 programs were approved,
both to do infrared direct imaging (GO-11702, PI = Yan; GO-11700 PI =
Trenti). Pure-parallel observing is perfectly suited to the goals of
the WISP Survey, since it allows us to obtain deep, continuous, IR
spectroscopy in dozens of uncorrelated fields.

We selected our parallel targets with a preference for long
integration times and high-galactic-latitude ($>30$\ degrees out of
the galactic plane) fields.  In order to reach similar depths in both
the G102 and G141 spectra, we planned an $\sim$2.5:1 G102:G141
integration time ratio.  Additionally, the process of extracting the
slitless spectra from the grism images requires direct imaging of the
same fields (see Section \ref{sec:datared}) and so we obtained imaging
in F110W and F140W (corresponding to J\textunderscore110- and H\textunderscore140-band, respectively) on
the same orbits as the grism imaging with an $\sim$6:1 grism:direct
imaging integration time ratio.  To date, we have received, reduced,
and analyzed 19 fields (Table~\ref{tab:schedule}), the results of
which are presented here.

Cycle 17 was the first year a powerful new procedure was adopted for
scheduling and optimizing parallel observations. Unlike previous
cycles, each of the Parallels teams was allowed to select individual
observing opportunities, with accurately known durations of each
orbital visit, well in advance.  Since every parallel visit is unique,
with its own combination of full- and partial orbits, we optimized each individual orbit to obtain the exposure time ratios mentioned above in the available time, thereby reaching a carefully balanced set of sensitivities. This tremendous improvement over
earlier years of ``generic'' parallel visits has improved the
efficiency of our observing program by about 50\% for a fixed
allocation of orbits. We also requested, and were granted, an
additional 17 single-orbit parallel observations. Although these were
too short for the purposes of other programs, when we observe these
new parallel fields in only the F140W and G141, we are still able to
reach comparable depths to the G141 images of many of our multi-orbit
fields (See Table \ref{tab:schedule}).


  
\section{Data Reduction}\label{sec:datared}
Slitless spectroscopy offers a considerable number of advantages for
our WISP program. The high multiplexing capabilities offered by the
WFC3 grisms combined with the observation of a significant number of
uncorrelated fields offered by the pure parallel strategy, makes it
perfectly suited for survey purposes. It allows a wide, uncontaminated spectral coverage with stable and accurate calibration. The ability to combine G102 and G141 grisms, with a well-behaved overlap gives the unique opportunity to find multiple lines per object at faint magnitudes. Because of the well-behaved photometry, it is possible to combine the spectra with broad-band photometry to provide comprehensive and precise spectral energy distribution (SED). Moreover, the use of the HST
offers a much lower sky background level compared to the
ground. However, to fully exploit the potential of this tool, one
needs to overcome some limitations. In crowded fields, object spectra
and orders sometimes overlap, leading to confusion in the spectral
extraction and to uncertainties in the background estimate. Given the
high latitude of the WISP fields, crowding did not represent a problem
in our observations. The spectral resolution is
determined by the size of the target: for point-like sources, this
corresponds to the instrument Point Spread Function (PSF), while for
extended sources it is the spatial extent of the object in the
dispersion direction.  

This basic principle of the slitless reduction is the use of the direct images to provide the input catalog of objects. The object positions, sizes and shapes are then used - in conjunction with a precise knowledge of the mapping between the direct and the dispersed frames to extract photometric and wavelength calibrated 2D and 1D spectra from the data using a 3D flat-field cube which contains the calibration of each detector pixel for each wavelength. This process has been integrated with the WFC3 pipeline (CALWF3 version 2.0) with the provision of the aXe software \citep{kuemmel09a} which has been developed to handle all of the HST slitless spectroscopic modes. aXe maps not only the + first order spectra but also the zeroth, - first, +/- second and higher order spectra which allows the quantitative estimation of contamination in the case of overlapping objects.

We processed all the data (including both grism and direct imaging
exposures) using the CALWF3 pipeline (version 2.0).  Direct images
were corrected for bias, dark, flat-field and gain variation using the
best reference files obtained from the STScI archive. WFC3 performs multiple non-destructive reads in the slitless spectra in MULTIACCUM mode with the sampling sequence SPARS100. This allows cosmic ray rejection in a single exposure sequence. The reduction pipeline fits the accumulating signal
in the the MultiAccum readouts to identify the cosmic rays.  
In the slitless grism data, pixels cannot be associated with a unique
wavelength. For this reason, CALWF3 applies a unity flat so that no
pixel-to-pixel correction is applied at this stage.  The proper
flat-fielding is then applied during the extraction of the spectra with aXe.

\subsection{Direct Images} 
Because our pure-parallel observations do not allow dithering, the
processed images still contain a significant number of bad pixels that
were not identified during the CALWF3 step. In addition, CALWF3 fails
to properly flag a fraction of cosmic ray hits, when these were
impacting multiple pixels. We identify and flag these pixels in the
data quality (DQ) masks using custom IDL routines. Finally, the images
are combined with the IRAF task MULTIDRIZZLE \citep{koekemoer02} using standard parameters. This
step also corrects for bad pixels and cosmic rays, and removes the
geometric distortion using the latest distortion solution (IDCTAB)
files. The final drizzled images have accurate astrometry and a
uniform photometry across the image.

The position of the spectra in the dispersed images is determined by
the location of the objects in the corresponding direct exposure.  We
run SExtractor \citep{bertin96} on the drizzled direct images,
creating independent source catalogs for the F110W and F140W images.
For the source detection we used the RMS weight maps derived during
the MULTIDRIZZLE step. Every source in the catalogue is required to
have at least 6 connected pixels (DETECT\_MINAREA $= 6$), with values
above 3.5$\sigma$ (DETECT\_THRESH $=3.5\sigma$). Finally, the catalogs
generated for the F110W and F140W images are matched and put on the
same numbering scheme. The individual detection in both filters allow us to keep objects detected in only one band and ensures an accurate astrometry between the grism image and the reference direct one. Because aXe performs the
spectral extraction on individual (not distortion-corrected) grism
frames, the combined catalog is projected back to these frames to generate an input
object list (IOL) of coordinates for each grism frame to be used in
the aXe extraction. Starting from these IOLs, the aXe software
computes the reference position and spectral trace for each object in
the dispersed images.
 
\subsection{Grism Images} 
Before proceeding to the actual spectral extraction, the sky
background must be estimated and subtracted from each grism frame. It
is possible, within aXe, to estimate the background locally by
interpolating on the sky regions above and below the spectral
trace. However, we decided to perform a global sky subtraction on the
entire image before proceeding to the extraction of the spectra. This
method proves more efficient for slitless survey data
\citep{pirzkal04}, where spectra may be be contaminated by the trace
of nearby objects, and multiple orders. 

In the slitless mode, each spatial element of the detector receives
the full sky and telescope foreground emission integrated over the
grism bandpass. The grisms disperse the foreground light, but the lack
of a slit results in a mapping of spatial to spectral domains that
reintegrates uniform sources (e.g., the sky) effectively into
undispersed emission. A portion of the detector does not receive the
full zero-order sky and thus there is a ramp in the sky in the first
one hundred or so columns on the detector. The sky foreground in the
G141 and G102 grisms is a combination of zodiacal light, earth limb
and thermal emission from the telescope. The processed two-dimensional
grism images contain a significant signal from these sources that must
be removed for proper photometric calibration, as well as for optimal
identification of spectral features.

We constructed a master sky template for each grism by making a stack
of all of our grism exposures.
The individual images were first scaled using the image mode, then
combined with a median stack. The resulting master sky contains
low-level residuals from bright sources, but these have little impact
on our results. Our sky template is being improved as additional
independent fields are observed and added to the master sky.

For each grism observation we scaled the appropriate master sky and
subtracted this from the post-pipeline two-dimensional image. We
experimented with scaling the sky by the mode of the pixel
distributions and by minimizing the sigma in the sky subtracted image.
These two approaches gave nearly identical results, and we proceeded
with the skies scaled using the mode. The scale factors varied by more
than a factor of two from one field to another, reflecting the
changing zodiacal light background and contribution from the Earth's
limb.

As in the direct images, the CALWF3 processed data still showed
residual bad pixels and cosmic rays.  We constructed a bad-pixel mask
in which we flagged all bad pixels, including those with low-QE and
with high value of the dark current.  We then interpolated over these
flagged pixels along the dispersion direction using a second order
polynomial. We find that the majority of the bad pixels are isolated,
and most interpolations were over at most $3$ pixels. Thus the
interpolation step does not introduce spurious features into the
spectra.  Moreover, because the position of each bad pixel is
recorded in the masks, we have checked that the identified emission
lines (see below) were not impacted by the interpolation
process. Finally, we created the final dispersed images by averaging
together the corrected frames, weighting each image by its exposure
time. These frames were used in our visual inspection.

\begin{figure*}[htbp]
   \centering
   \includegraphics[width=14cm]{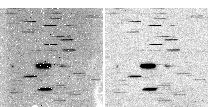}\\ 
   \vspace{0.3cm}
  \includegraphics[width=14.1cm]{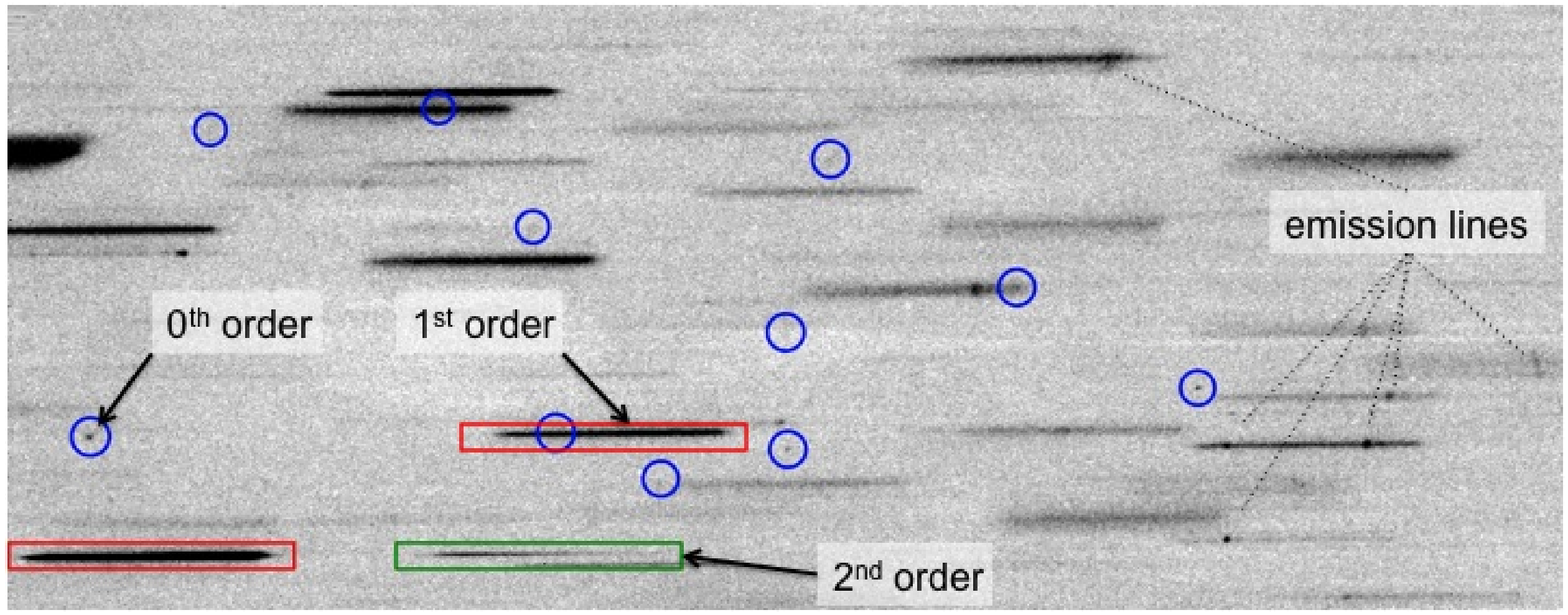}
  \caption{{\it Top panel} shows a comparison of an individual G141 grism image before and after the reduction process. The raw image on the left shows a large number of bad pixels and cosmetic artifacts that are no longer visible on the clean image. {\it Bottom panel:} An example of different spectral features in our grism images. Red and green rectangles indicate examples of first and second order spectra respectively. The blue circles mark the zero order positions,  indicating in some cases a contamination to the spectra. Some emission lines are also indicated.}
   \label{fig:raw_clean}
\end{figure*}

We also run MULTIDRIZZLE on the individual grism frames. This has the advantage of performing cosmic ray rejection and flagging the affected pixels in these frames. However the drizzled grism image is not used for extraction. The aXe procedure extracts the spectra from the individual grism images and all flagged pixels will be ignored during this process. Another advantage of applying MULTIDRIZZLE on both direct and grism frames is to avoid potential differences in the WCS data for the two images that could lead to a mismatch in the source positions in the direct and grism individual images during the extraction.
\subsection{aXe Extraction}

The spectral extraction is performed with the slitless extraction package aXe (version 2.0) developed at Space
Telescope European Coordinating Facility (ST-ECF). 
The input image list is prepared with each individual grism image paired with the direct image which is closest in time and its SExtractor catalog. 
We used the latest in-orbit configuration files \citep{kuntshner09a, kuntshner09b} and thermal vacuum (TV) 3 sensitivity calibrations. Some of these reference files will be improved for later reductions.
We compared several apertures and orientation parameter setups to optimize the extraction. 
We used an aperture width of twice the object size listed in the input object list, and an extraction direction 
perpendicular to the spectral trace (see aXe user manual\footnote{http://www.stecf.org/software/slitless$\_$software/axe/} for details).

Two-dimensional cutout spectra of each object in each exposure are
extracted and then combined (with aXeDRIZZLE) to create a final high
signal-to-noise ($S/N$) ratio spectrum. The combined 2D spectrum is
then used to extract the final object spectrum. We used optimal
extraction to increase the $S/N$ ratio of the
extracted spectra. The spectral contamination by spectra of nearby
objects is estimated by aXe using the object positions and magnitudes
in the direct image. We adopted the Gaussian emission model, using
SExtractor size parameters, to approximate the object morphology. The
simulated spectrum is then extracted and used to estimate the
contamination to each object.

\subsection{Emission line Objects}
Each stacked two-dimensional grism frame was searched visually by two
or more of the authors. We used the dispersion solution from the WFC3
configuration files to determine the positions of the aperture beams in
both grisms. The 1st order of each object is then marked with the
corresponding aperture beam and the original ID number in the IOL. The actual search area is about 3.3 arcmin$^{2}$, smaller than the IR channel field of view. Because of the shift between direct and grism images, spectra on the left part of the field do not have detections in the direct image, and zeroth orders cannot be identified at the far-right side of the frame. Some examples of 1st and 2nd order aperture beams are shown in
Figure~\ref{fig:raw_clean} by the long rectangular boxes.

\begin{figure*}
\centering
\vspace{-1.3cm}
\includegraphics[width=12cm]{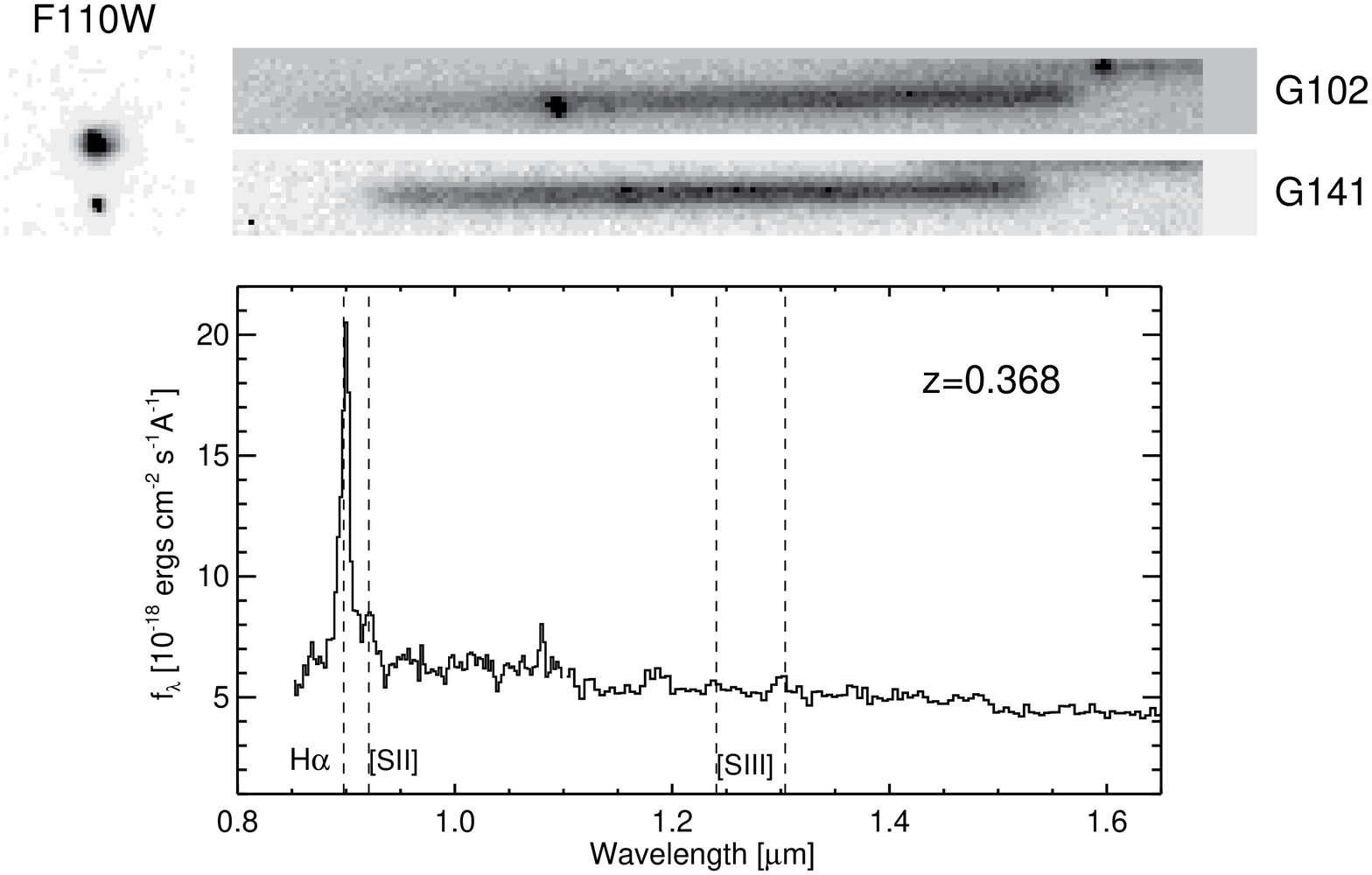}\\
\vspace{-1.3cm}
\includegraphics[width=12cm]{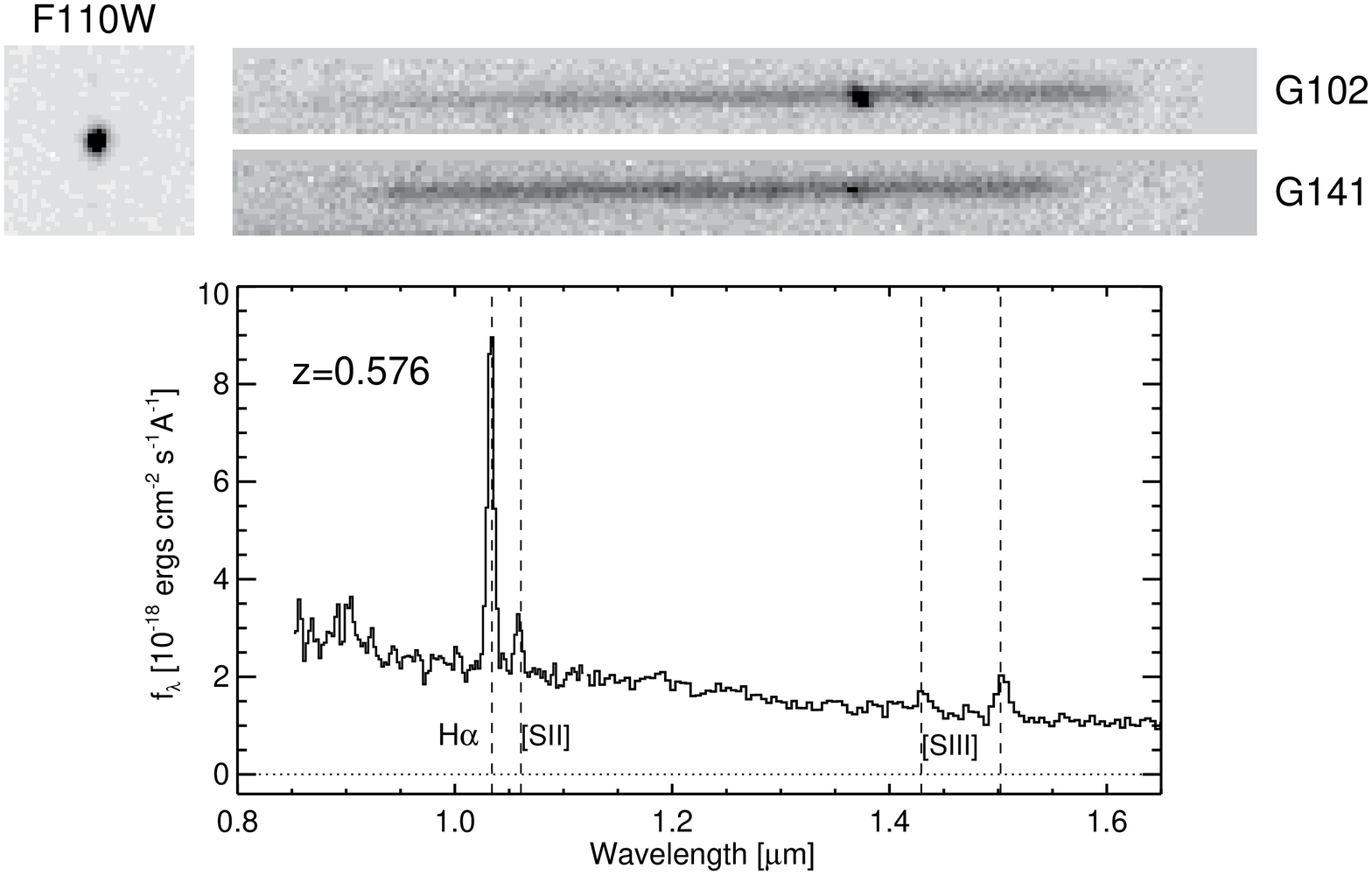}\\
\vspace{-1.3cm}
\includegraphics[width=12cm]{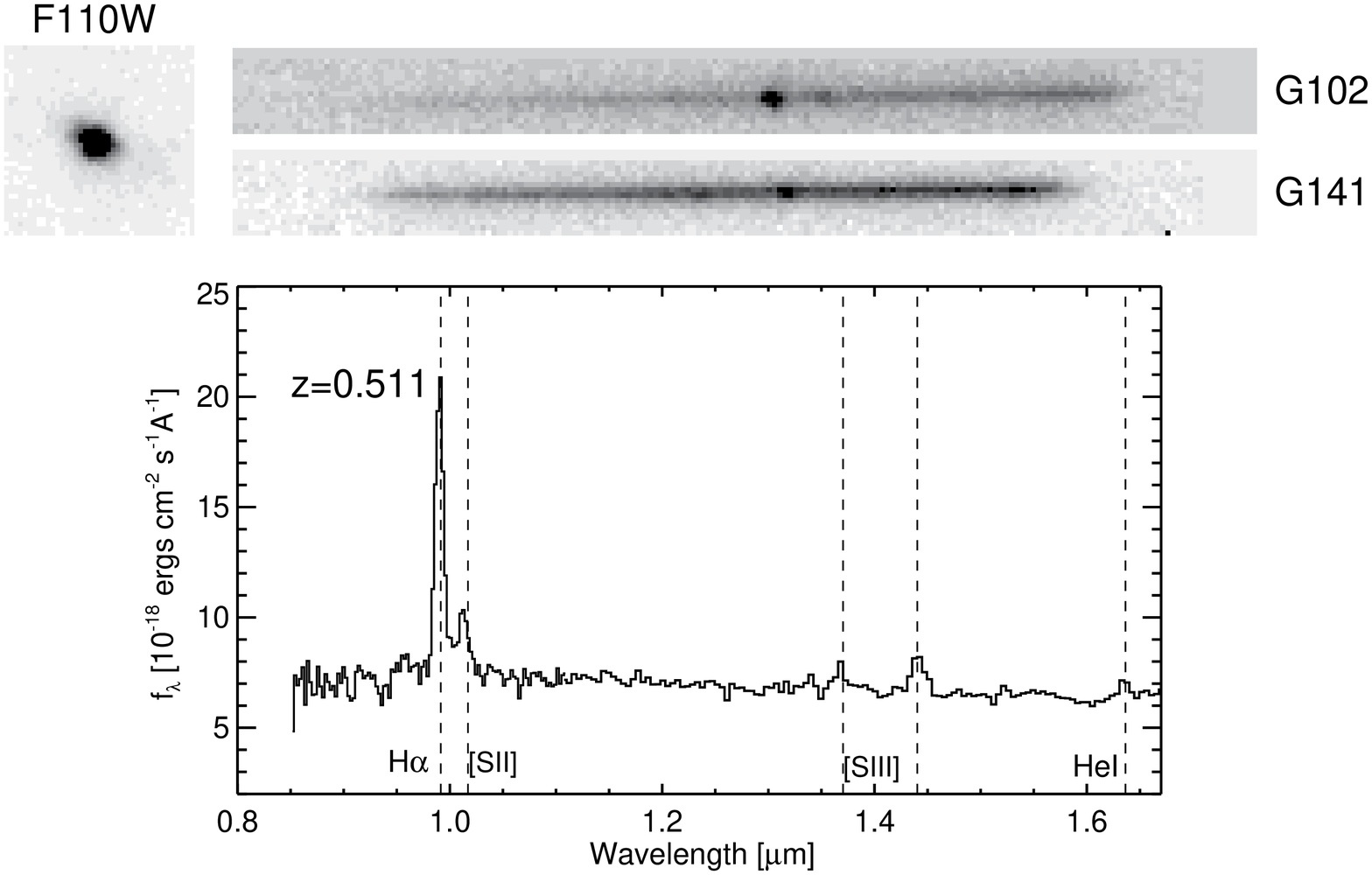}
\vspace{-0.5cm}
\caption{Examples {\sc I}: low-redshift emission line objects at $ 0.4 < z < 0.8 $. For each object we show the direct image cut-out ($5\arcsec \times 5\arcsec$), the 2D G102 and G141 grism spectra and the 1D extracted spectrum at observed wavelength. In this category, \ha\ falls in the G102 grism. We can see several emission lines: [\ion{S}{2}]$\lambda\lambda$6716+6732 \AA, [\ion{S}{3}]$\lambda$9069 \AA, [\ion{S}{3}]$\lambda$9532 \AA\ and \ion{He}{1}$\ \lambda$10830 \AA. }
\label{fig:examples1}
\end{figure*}

\begin{figure*}
\centering
\vspace{-1.3cm}
\includegraphics[width=12cm]{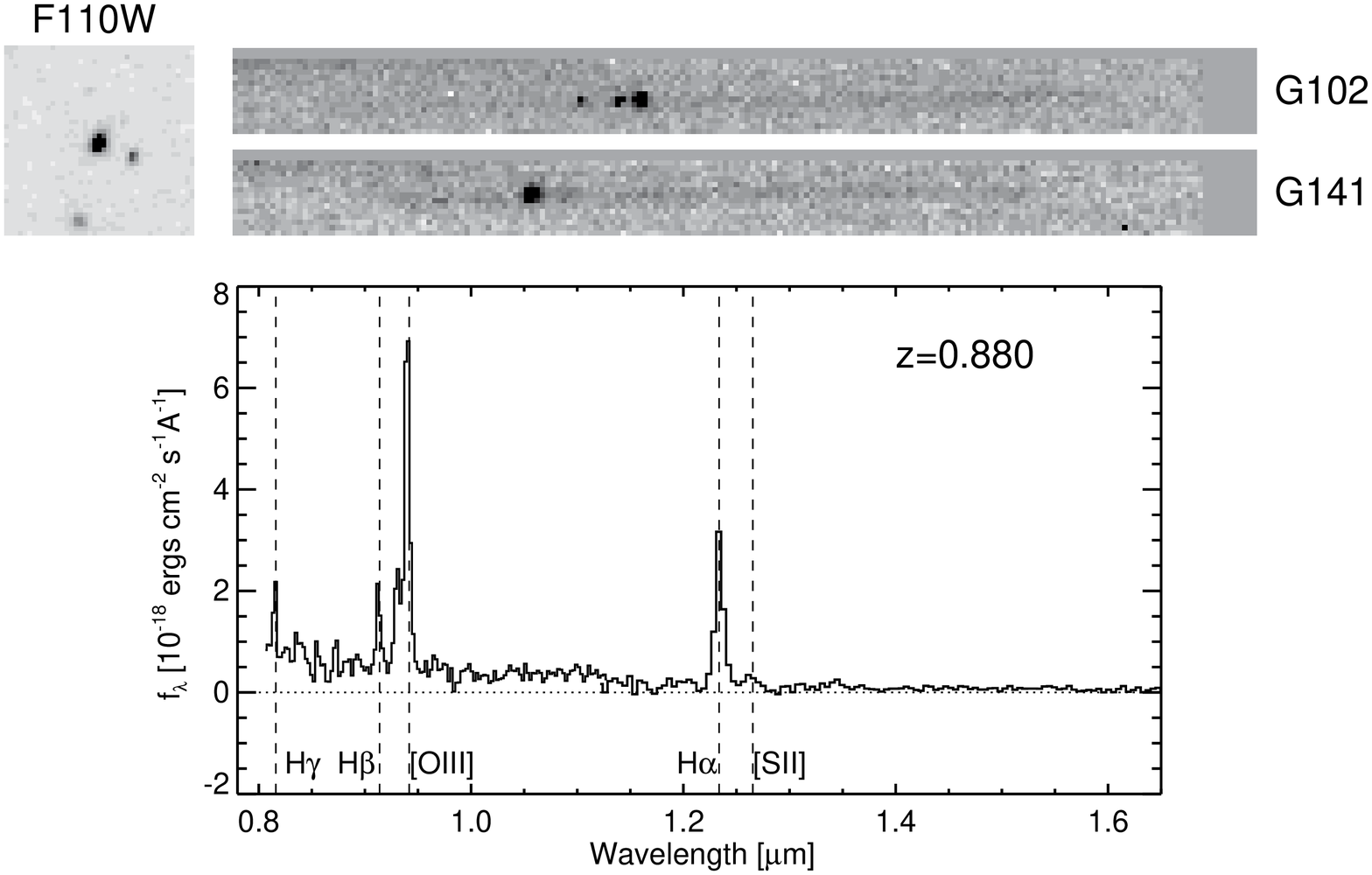}\\
\vspace{-1.3cm}
\includegraphics[width=12cm]{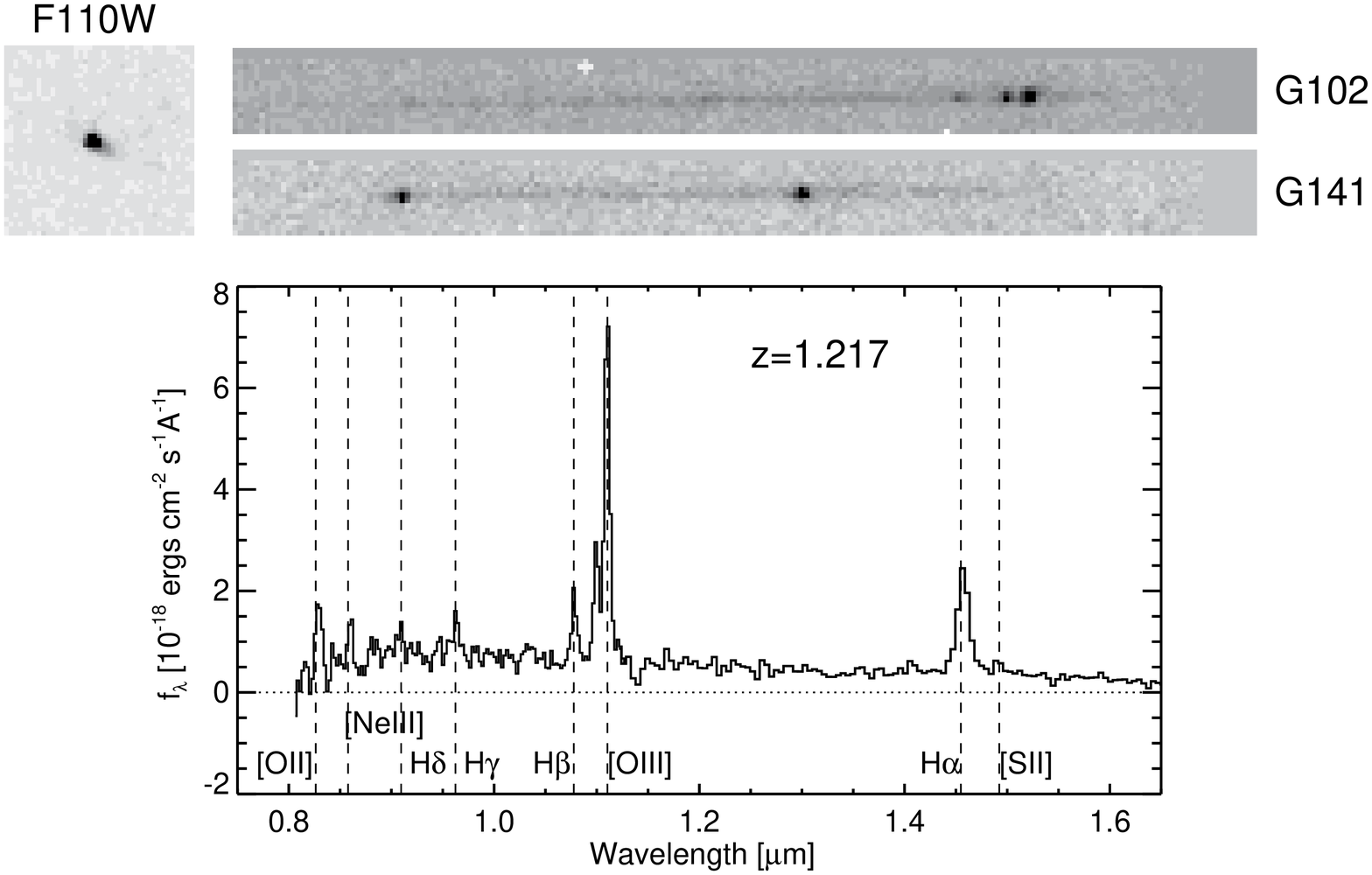}\\
\vspace{-0.5cm}
\caption{Examples {\sc II}: emission line objects in the redshift range $ 0.6 < z < 1.3 $. For each object we show the direct image cut-out ($5\arcsec \times 5\arcsec$), the 2D G102 and G141 grism spectra and the 1D extracted spectrum at observed wavelength. This category includes mainly \ha\/[\ion{O}{3}] line emitters, but we also observe [\ion{O}{2}], [\ion{Ne}{3}], \hb, \hg\ and [\ion{S}{2}]$\lambda\lambda$6716+6732 \AA.}
\label{fig:examples2}
\end{figure*}

\begin{figure*}
\centering
\vspace{-1.3cm}
\includegraphics[width=12cm]{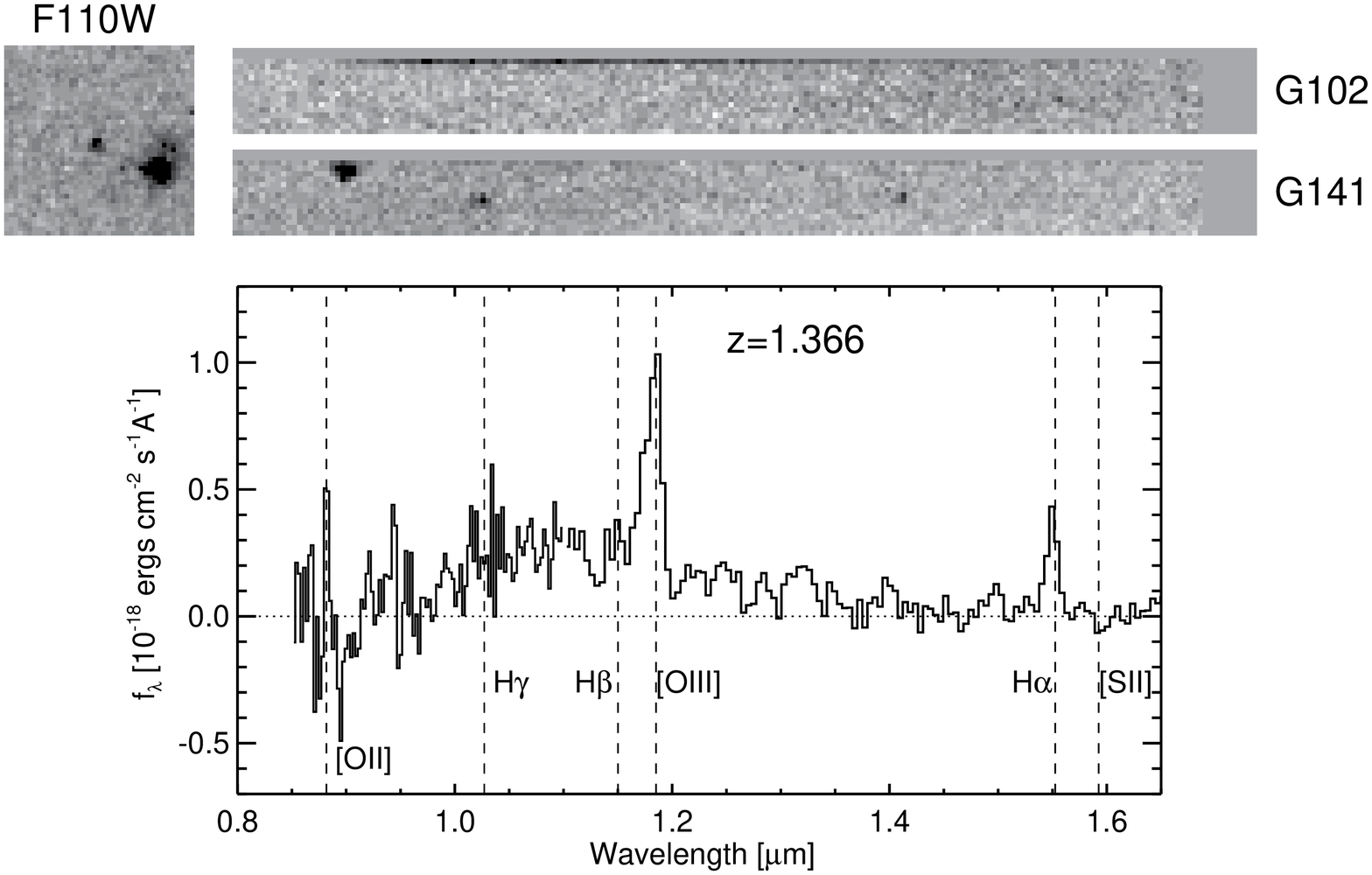}\\
\vspace{-1.3cm}
\includegraphics[width=12cm]{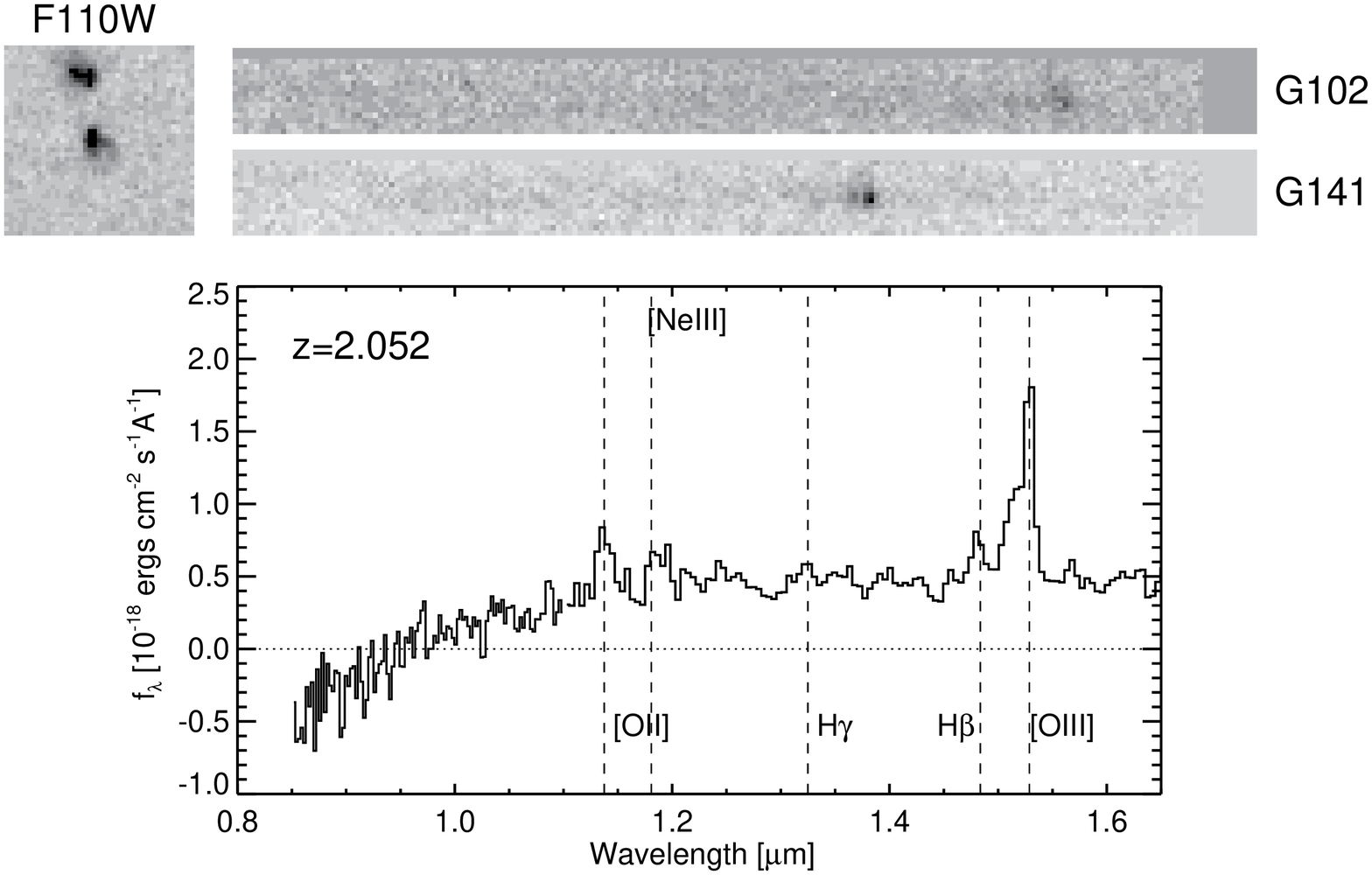}\\
\vspace{-1.3cm}
\includegraphics[width=12cm]{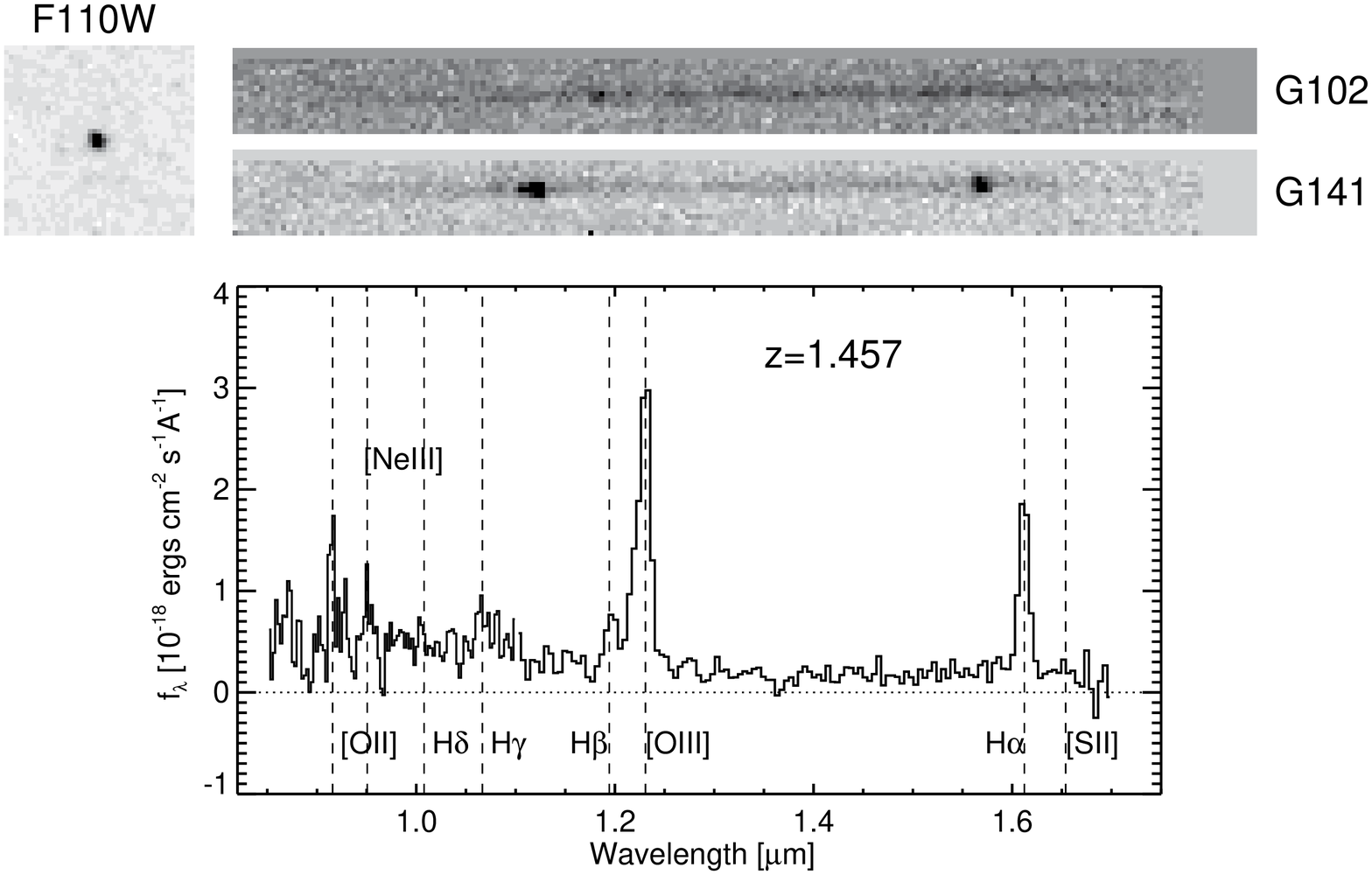}\\
\vspace{-0.5cm}
\caption{Examples {\sc III}: high-redshift emission line objects between $1.3 < z < 2.3 $. For each object we show the direct image cut-out ($5\arcsec \times 5\arcsec$), the 2D G102 and G141 grism spectra and the 1D extracted spectrum at observed wavelength. At these redshifts, both \ha\ and [\ion{O}{3}] are shifted to the G141 grism. In many cases we also observe the [\ion{O}{2}] emission line in the G102 grism.}
\label{fig:examples3}
\end{figure*}

Besides the overlap between the objects, the most important spurious
feature is the zero-order image. Because the light is not dispersed,
the zero-order image could be easily confused with an emission line.
We derived the position of all the
zero-orders in the grism exposures, using the positions of all the
sources in the direct images. We marked the zero-orders in the grism
exposures to be able to exclude them during the visual inspection of
the grism images. In Figure~\ref{fig:raw_clean} we show an example of
this procedure: the blue circles show the zero order location of some of the
objects detected in the direct images.
With all possible contaminants identified in the grism images, we
proceeded to compile a list of all emission-line source in each field.
We compared the G102 and G141 spectra of each field side-by-side, to
aid in the identification of genuine emission-lines. In many cases the
identification of the lines is
unambiguous. 
For exampe, galaxies with H$\alpha$ often also have detectable
[SII]6717/6731 and sometimes [SIII]9069,9532 emission (Figure \ref{fig:examples1}). Strong
[OIII]5007,4959 emission in the G102 grism is easily recognized on the
basis of H$\alpha$ in the G141 and often H$\beta$ in the G102 spectra,
even though the two [OIII] lines are blended (see Figure \ref{fig:examples2}). At higher
redshifts, strong [OIII]5007,4959 in the G141 grism is
confirmed by faint H$\beta$ emission. In some cases [OII]3727 can be seen in the G102
spectra for sources with strong [OIII]5007 in the G141 spectra (Figure \ref{fig:examples3}).

There are a large number of objects with only a single faint emission-line in 
the full spectral range covered by the two grisms. It it most probable that these
are weak H$\alpha$ emission lines for which we are unable to detect other lines. However, since we see many [OIII]/\ha\ pairs up to $z \sim 1.5$ we expect, in absence of other indication, to have a fraction of the single lines in the G141 that would correspond to [OIII] whereas \ha\ has moved out of the G141 grism. The [OII]3727 line is seldom recognized as a redshift indicator by
itself, since in this case it is more likely to be an \ha\ line at low-z rather than a high-z [\ion{O}{2}] line.
There are some objects, however, for which we believe the single line in the G141
spectrum is [OIII]5007,4959. This is primarily based on the lack of continuum
in the G102 spectra, indicative of the fall off in the continuum below the 
Balmer and 4000 angstrom breaks. 

Each emission feature was assigned a confidence flag, ranging from 0 to
4 in order of decreasing confidence. The highest
confidence level (class 0) was assigned to objects with multiple emission features and
unambiguous redshifts. Confidence classes 1, 2 and 3 were assigned to single features
with unambiguous identifications (class 1), good quality (class 2), moderate quality (class 3), and uncertain (class 4, which are not included in the present analysis).  
 Similar confidence classes have been used in
ground-based redshift surveys (e.g. CFRS, \citet{lefevre05}; GDDS, \citet{abraham04}) and
have been found to be useful in restricting various types of analysis to high confidence
objects. Monte Carlo simulations will be used to assess the flux completeness and 
reassess the confidence levels when the data set is complete.

 
The final object catalog contains a measurement of each emission line,
performed using a custom IDL fitting tool, and relevant information
from the object extraction. The line fitting uses the pixel position
of the identified lines, applies the grism dispersion solution to find
the corresponding wavelength and therefore the redshift of the object.
The fitting model consists of a polynomial continuum and a gaussian
(a multi-gaussian model is used to deblend the \ha-[S{\sc ii}] and
[O{\sc iii}]$\lambda$5007,4959 \AA\ lines). The program fits the
identified lines by least squares, and tries to fit potential lines
that could be present in the spectrum at the correct redshift with a
lower significance of typically 2 $\sigma$. The entire process is
visually monitored by displaying the line, the fit model and the
robustness of the fit. The fitting flux uncertainties are derived from
our aXe extraction and calibration. As an independent check, a large
subset of these same emission lines were also measured by hand, using
the SPLOT task in IRAF. The resulting splot measurements generally
agree with the automated measures to within the uncertainties.

We investigated the
accuracy of the grism data calibration. First, we compared the fluxes
of lines detected simultaneously in both G102 and G141 grisms, in a
wavelength domain restricted to a throughput higher than 10\%. In most
cases the flux agrees to better than 10\% between the two grisms.
Second, we compared the agreement between wavelength calibrations by
using the overlapping region around 1.15 micron, and the redshift
determined independently from different lines detected in G102 and G141. We
observed a typical offset of 2 $\times 10^{-3}$ microns that is within the
uncertainties of the redshift determination in the individual grisms.


\begin{figure}
  \centering
\includegraphics[width=8cm]{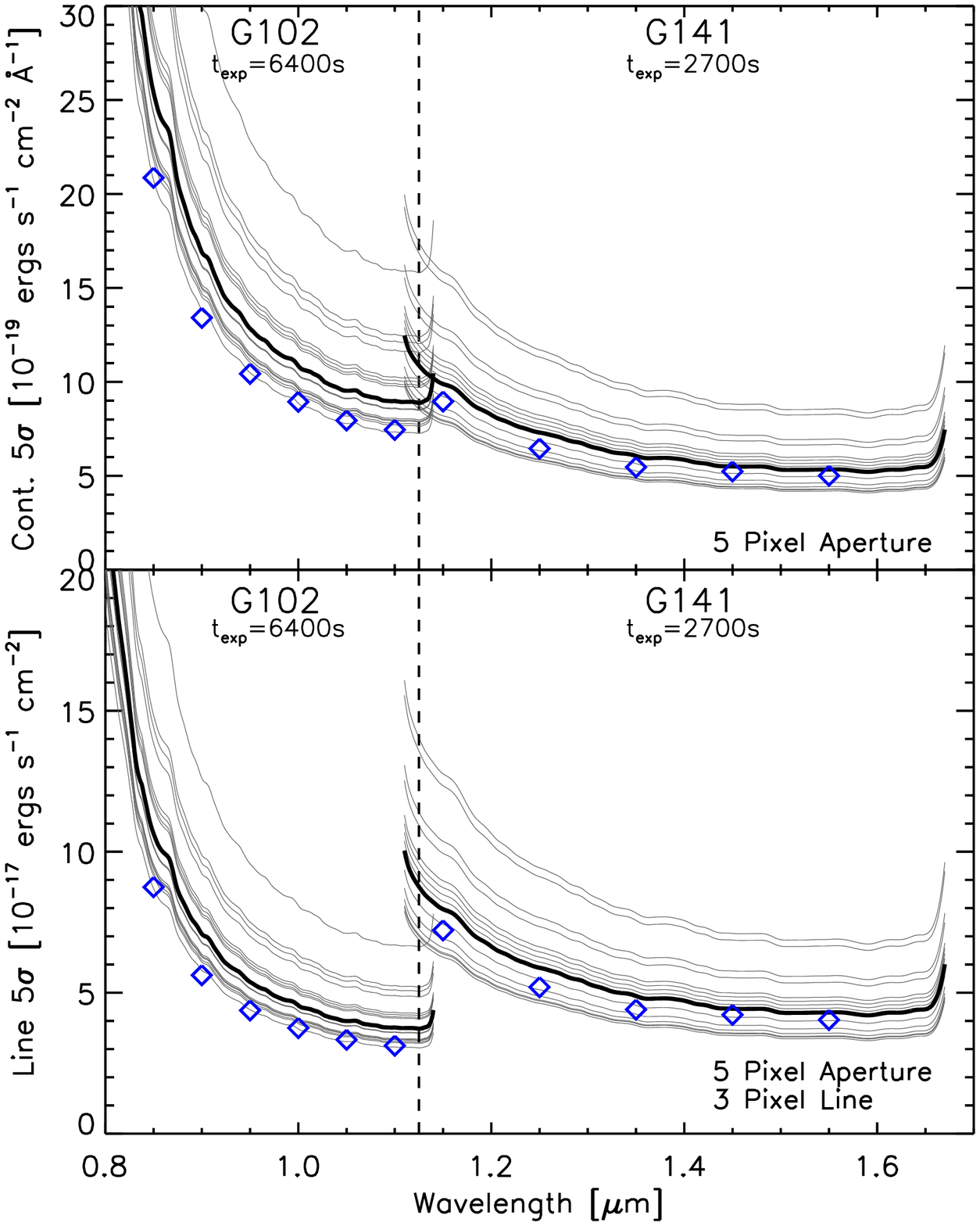}
\caption{The $5\sigma$ continuum (top) and line (bottom) sensitivities as a function of wavelength. The depth of each field is plotted in gray and scaled (by exptime$^{1/2}$) to the ``nominal'' survey exposure times. The thick solid lines denote the median field depth. The blue diamonds are the values given by the exposure time calculator assuming ``average'' earthshine and zodiacal light backgrounds. A five-pixel (0.65$''$) wide aperture was used for these estimates and the line sensitivities assumed three pixel widths in the dispersion direction. More compact sources will give slightly better sensitivities. There is significant scatter from the median due to variations in background levels.}
\label{fig:sens}
\end{figure}

\begin{figure}
  \centering
\includegraphics[width=8.5cm]{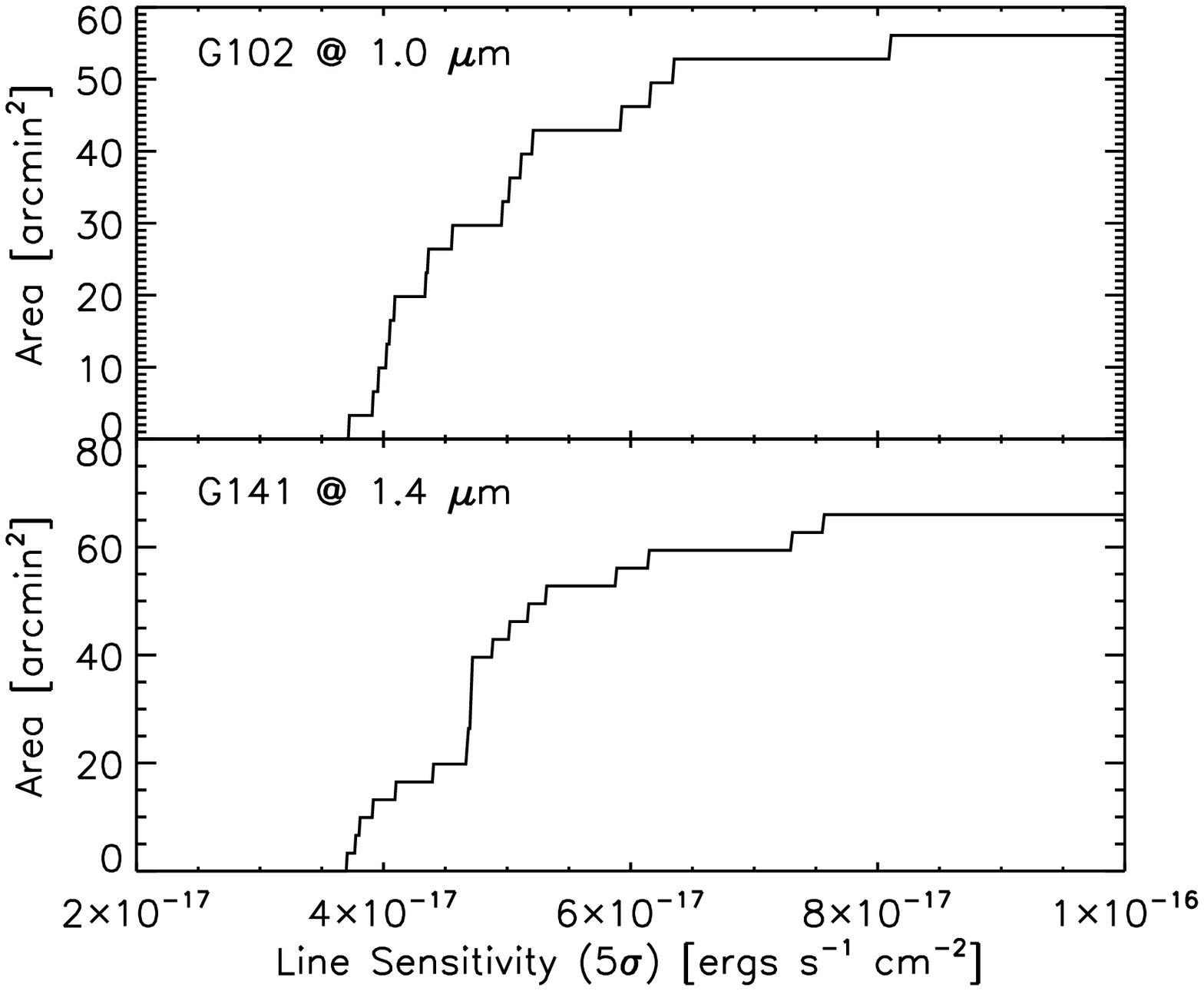}
\caption{Cumulative area as a function of depth.  Each field has an effective area of 3.3 arcmin$^2$ which is added to the cumulative area at the value of its five sigma line sensitivity (in a 5x3 pixel region).  The depth varies as a function of wavelength so we choose to use the depths at the central wavelengths (1.0 $\mu$m for G102 and 1.4 $\mu$m for G141).  The depths are better at longer wavelengths and worse at lower wavelengths.  The median depths in both grisms are $\sim5\times10^{-17}$ ergs s$^{-1}$ cm$^{-2}$.}
\label{fig:area_depth}
\end{figure}

\subsection{Grism Sensitivities}

Due to the pure parallel nature of our program, we have no control over
the location and timing of our observations.  Because of this, we are
subject to a variety of visit durations (typically 3-5 orbits), number of
readouts, and backgrounds (zodiacal light and earthshine), all of which
affect the sensitivity of our observations.  In order to determine the
depth of our spectroscopy, we have measured the background rms (in
electrons per second) of every final reduced grism image.  This rms is
then divided by the sensitivity (as a function of wavelength) and divided
again by the dispersion (in \AA\ pixel$^{-1}$) to give the $1\sigma$ noise
as a function of wavelength.  In Figure \ref{fig:sens} we plot the $5\sigma $
continuum sensitivity in a five-pixel wide aperture (in the spatial direction) for each pointing. We used three pixels width in the dispersion axis to estimate the line sensitivities. Each field has been scaled by (exptime) $^{1/2}$ so that the relative
depths are comparable. Though most observations agree with the predicted
depths from the WFC3 exposure time calculator, there is a large dispersion
in the depths due to background variations, and any given field can vary
substantially from the mean by typically 20\% (up to $\sim50$\%). We also plot in Figure \ref{fig:area_depth} our line sensitivity as a function of depth. In both grisms, we reach a median depth of about $\sim5\times10^{-17}$ ergs s$^{-1}$ cm$^{-2}$.  

In summary, for the typical exposure times of the survey ($\sim 6400$ s in
G102, $\sim 2700$ s in G141) we are able to detect (at $5\sigma$) compact
lines with $f = 5 \times$ 10$^{-17}$ ergs s$^{-1}$ cm$^{-2}$ at $0.95 < \lambda
< 1.17$ \AA\ and $1.3<\lambda<1.65$\AA.




\section{Emission Line Results}\label{sec:results}

To date, we have selected emission lines over 19 WFC3 fields, and
measured redshifts and line fluxes from the highest confidence
emitters.  While incomplete at the faintest lines, we clearly reach
emission line fluxes down to a few $\times 10^{-17}$ erg s$^{-1}$
cm$^{-2}$, often in the presence of a second brighter line.  
At the highest confidence levels, our preliminary analysis yields 328
(S/N $> 5$) emission lines from 229 objects in our 19 WFC3 fields. 
Simulations will be required to determine our
incompleteness due to extended emission, confusion, both high and low
equivalent widths, while accounting for the non-uniform survey
sensitivity shown in Figure \ref{fig:sens}. We list the identifications of these lines in Table 2. We note that the fraction of sources with multiple emission lines increases if we include lines detected at 3$\sigma$ to 5$\sigma$.


\begin{figure*}[!ht]
\includegraphics[width=8.5cm]{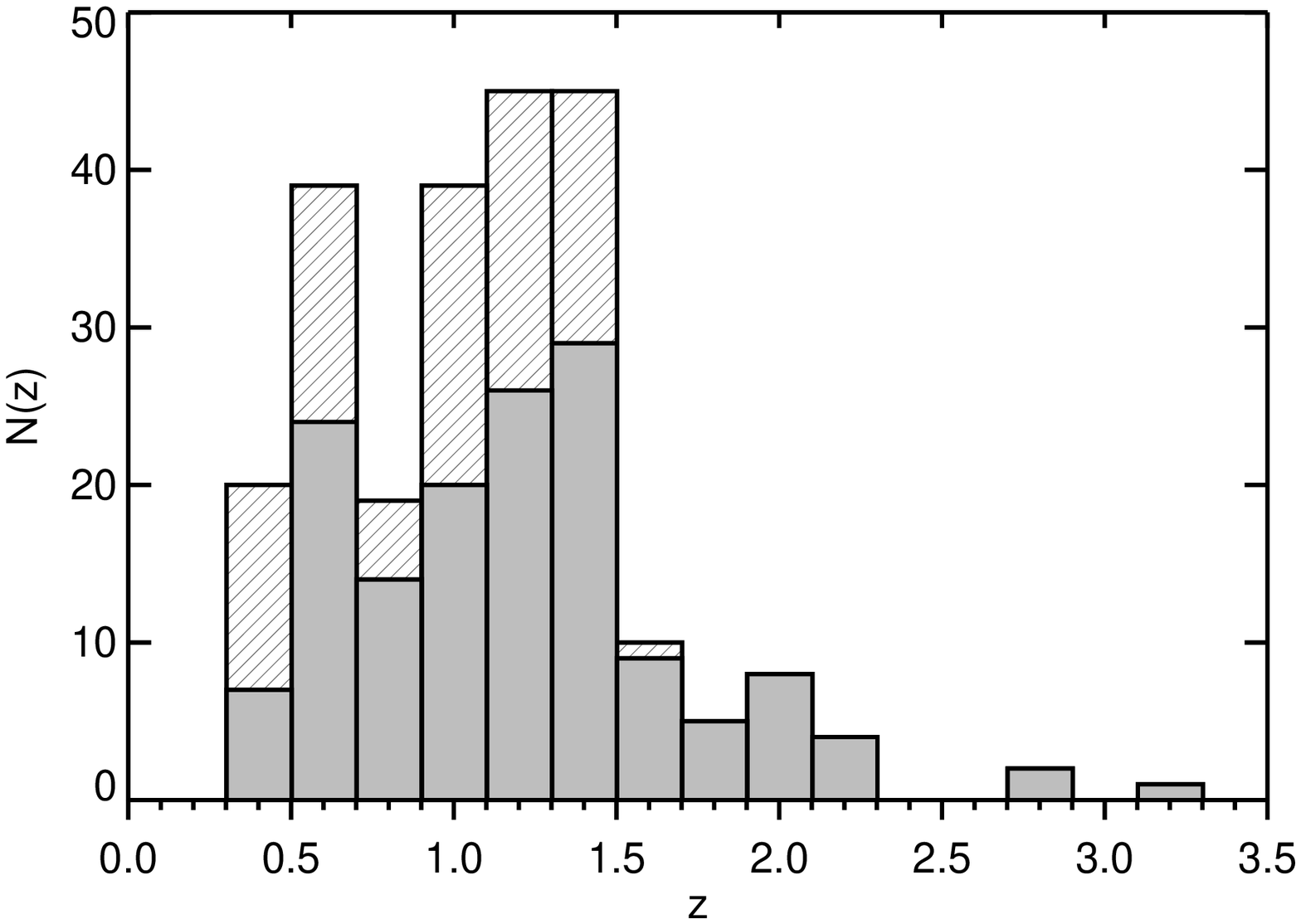}
\includegraphics[width=8.5cm]{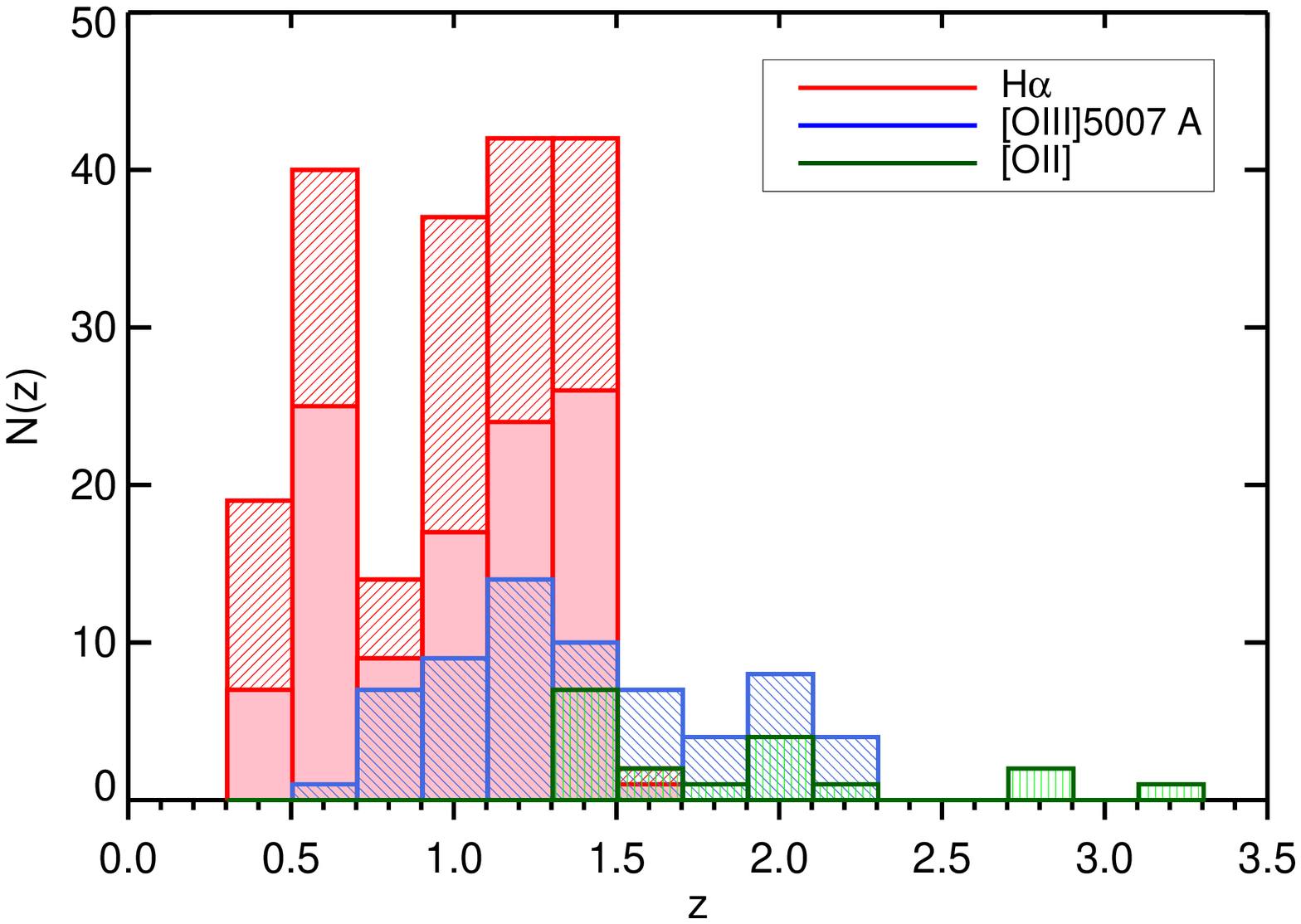}
\caption{ {\it Left --} The redshift distribution of our total sample of emission-line galaxies in 63 arcmin$^2$ of the WISP survey. Filled histogram represents our secure redshift determination (class 0 and 1), whereas on the top, the hatched part shows the uncertain single-emission-line objects assumed to be \ha\ emitters. {\it Right--} The redshift distribution of our galaxies according to their emission lines. Again the hatched part of the \ha\ emitters represent uncertain redshifts.}
\label{fig:zdist}
\end{figure*}

\begin{figure}[!ht]
\includegraphics[width=9cm]{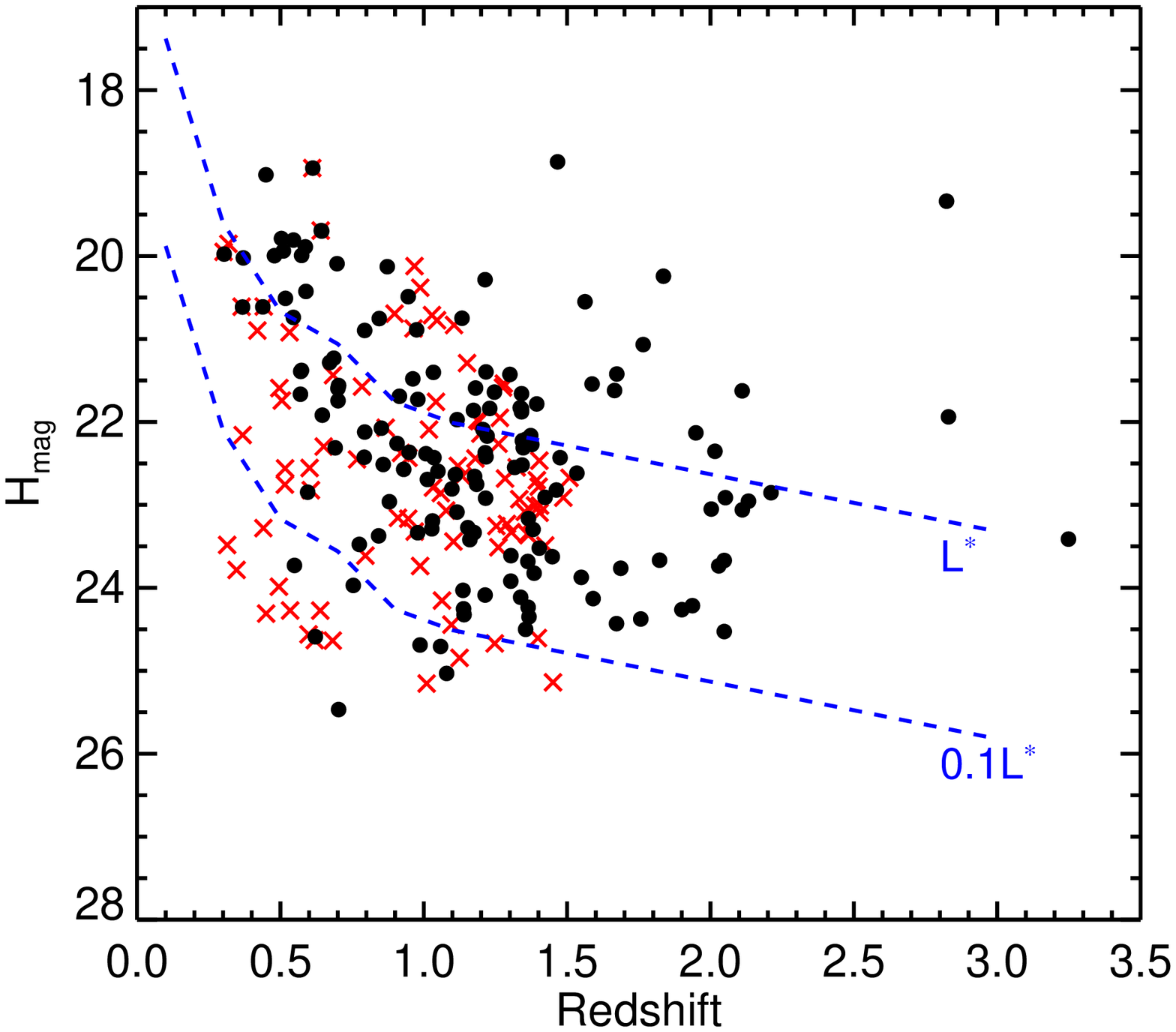}
\caption{H\textunderscore140-band magnitude as a function of redshift.  Secure redshifts are shown as black circles and single line (assumed to be \ha) emitters as red crosses.  We also show the apparent H\textunderscore140-band magnitude for L$^{\star}$ (and 0.1 L$^{\star}$) galaxies derived from rest-frame near-IR (at low-$z$) and optical (at $z > 0.7$) luminosity functions (Bell et al. 2003, Faber et al. 2007, Shapley et al. 2001).}
\label{fig:mag_z}
\end{figure} 

We find 185 galaxies with likely or definite \ha\ emission.
Amongst these, 56 have secure identifications based on multiple
emission lines.  We also identify single-line objects as
\ha\ emitters, as was the case with NICMOS grism spectroscopy
\citep{pmc99,hicks02,shim09}.  This assumption has generally worked in
ground-based narrow-band IR detections of single emission lines
\citep[e.g.][]{bunker95,mtm96,vanderwerf00,sorbal09}.

Other possible 
line identifications would almost always result in a brighter line that we would also detect. 
The most significant possibility of a single 
emission line that is not \ha, would be  [OIII] 5007+4959.  
This does occasionally happen in ground-based near-IR searches \citep{teplitz99}.
Not surprisingly, for the remaining 56 objects with multiple bright emission lines, 
the most common combination is \ha\ + [OIII] 4959/5007.  The redshift distributions for the total sample is shown in the left panel of Figure \ref{fig:zdist}, while the right-hand panel shows
the redshift distribution for the \ha, [OIII], and [OII] emitters separately.   

\begin{deluxetable} {l c}
\tablecolumns{2}
\tablewidth{0pt}
\tablecaption{Emission lines in the WISP Survey}
\tablehead{\colhead{Line} & \colhead{Number Detected}}
\startdata
Single lines & 129 \\
\ha\ confirmed & 56 \\
\hb\  & 5 \\ 
${\rm [OIII]~5007} $ & 59  \\
${\rm [OIII]~4959} $  & 35 \\
${\rm [OII]~3727} $  & 15 \\
${\rm [SII]~6717} $  + 6731  &15 \\
$ {\rm [SIII]~9069} $  & 3 \\
$ {\rm [SIII]~9532} $  & 5 \\

\enddata
\tablecomments{The main emission lines with a signal to noise ratio S/N $\geq$ 5 detected in the WISP Survey.}
\label{Line_tab}
\end{deluxetable}

\begin{figure*}
\centering
\vspace{-1.3cm}
\includegraphics[width=12cm]{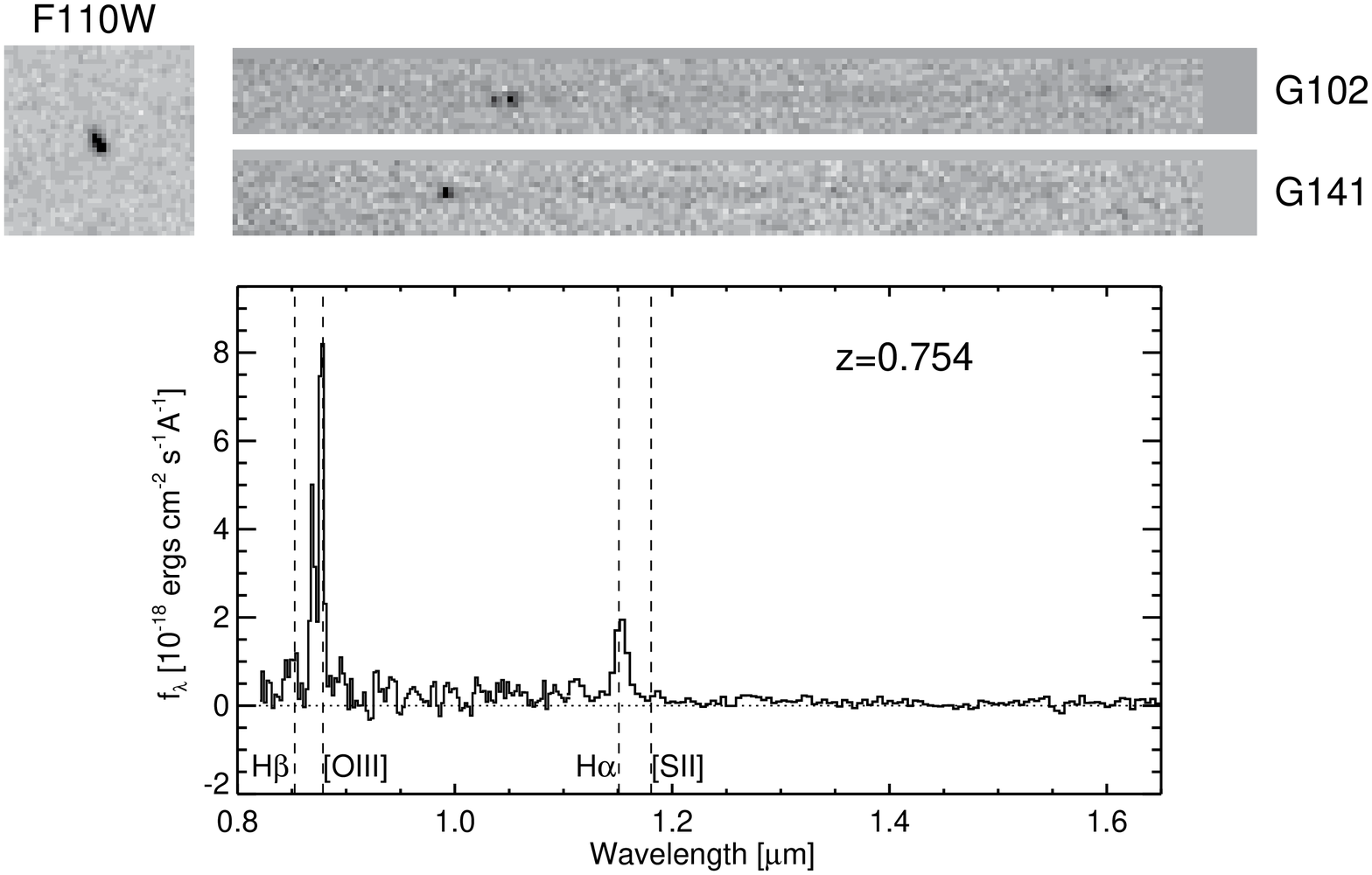}\\
\vspace{-1.3cm}
\includegraphics[width=12cm]{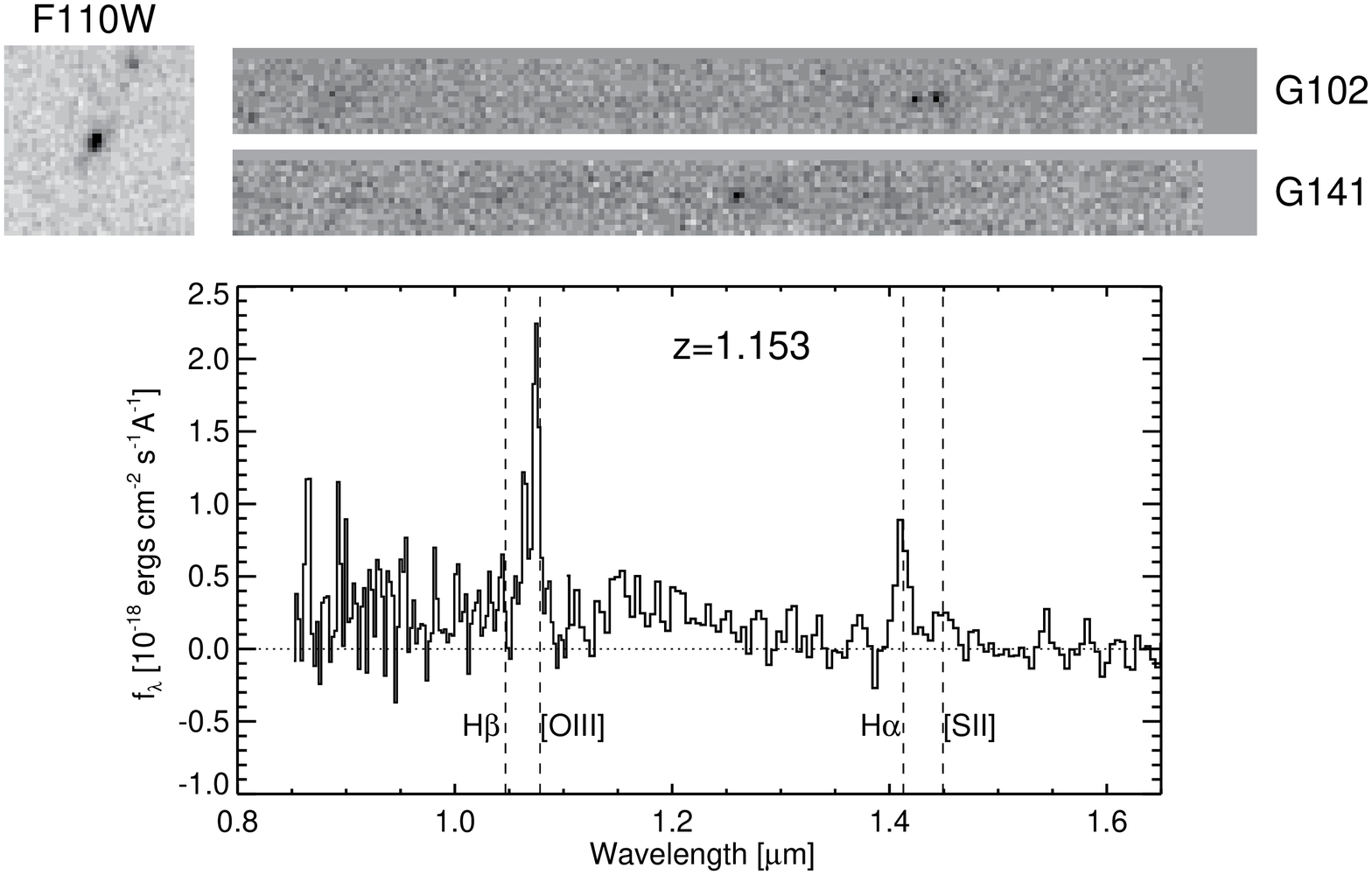}\\
\vspace{-1.3cm}
\includegraphics[width=12cm]{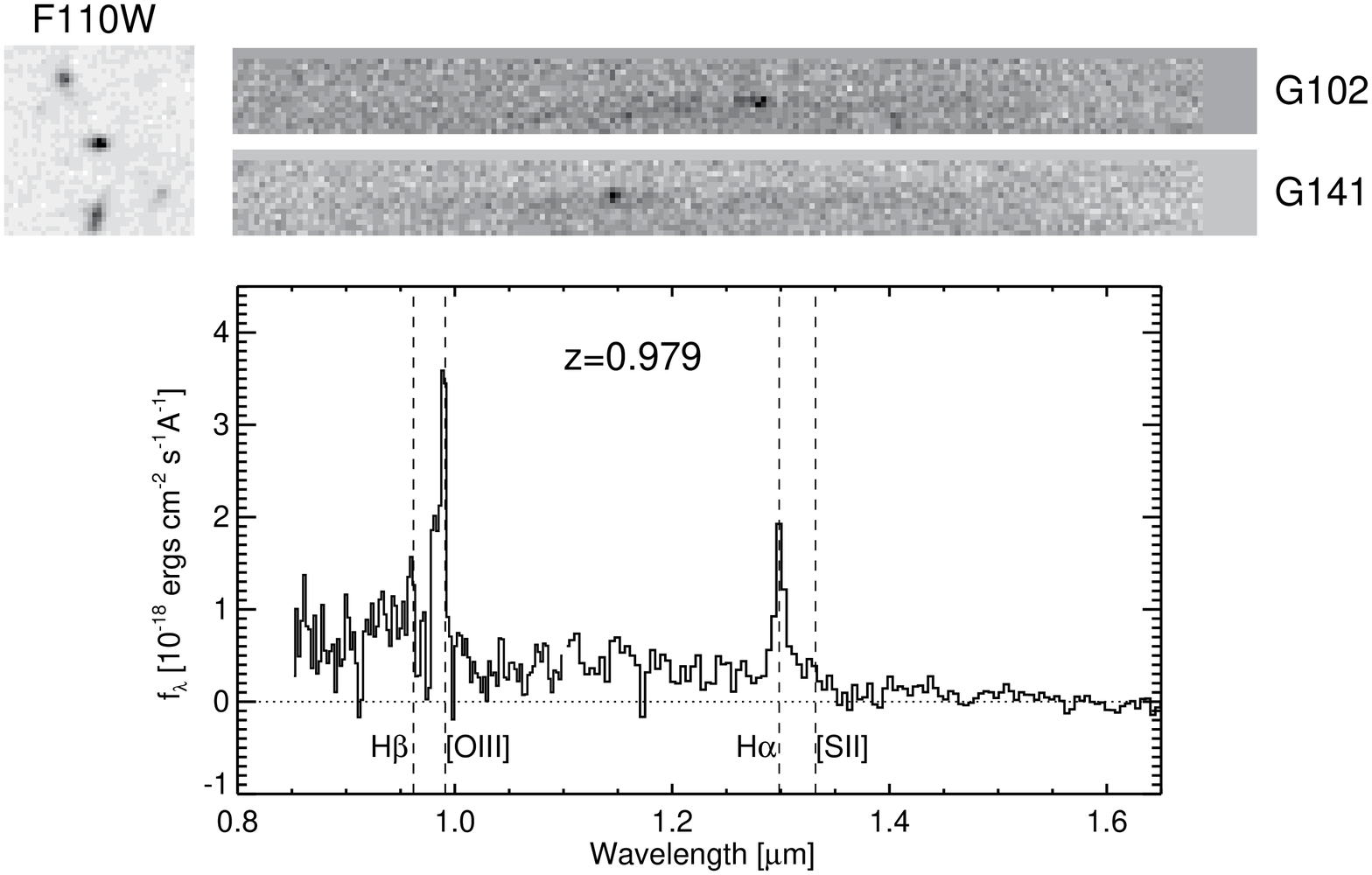}\\
\vspace{-0.5cm}
\caption{Examples {\sc IV}: very high equivalent width emission line objects. A strong emission line is observed in this group of galaxies with an extremely faint continuum which barely visible in the 2D grism images. For each object we show the direct image cut-out ($5\arcsec \times 5\arcsec$), the 2D G102 and G141 grism spectra and the 1D extracted spectrum at observed wavelength.}
\label{fig:examples4}
\end{figure*}

\begin{figure*}[!ht]
\centering
\includegraphics[width=8.5cm]{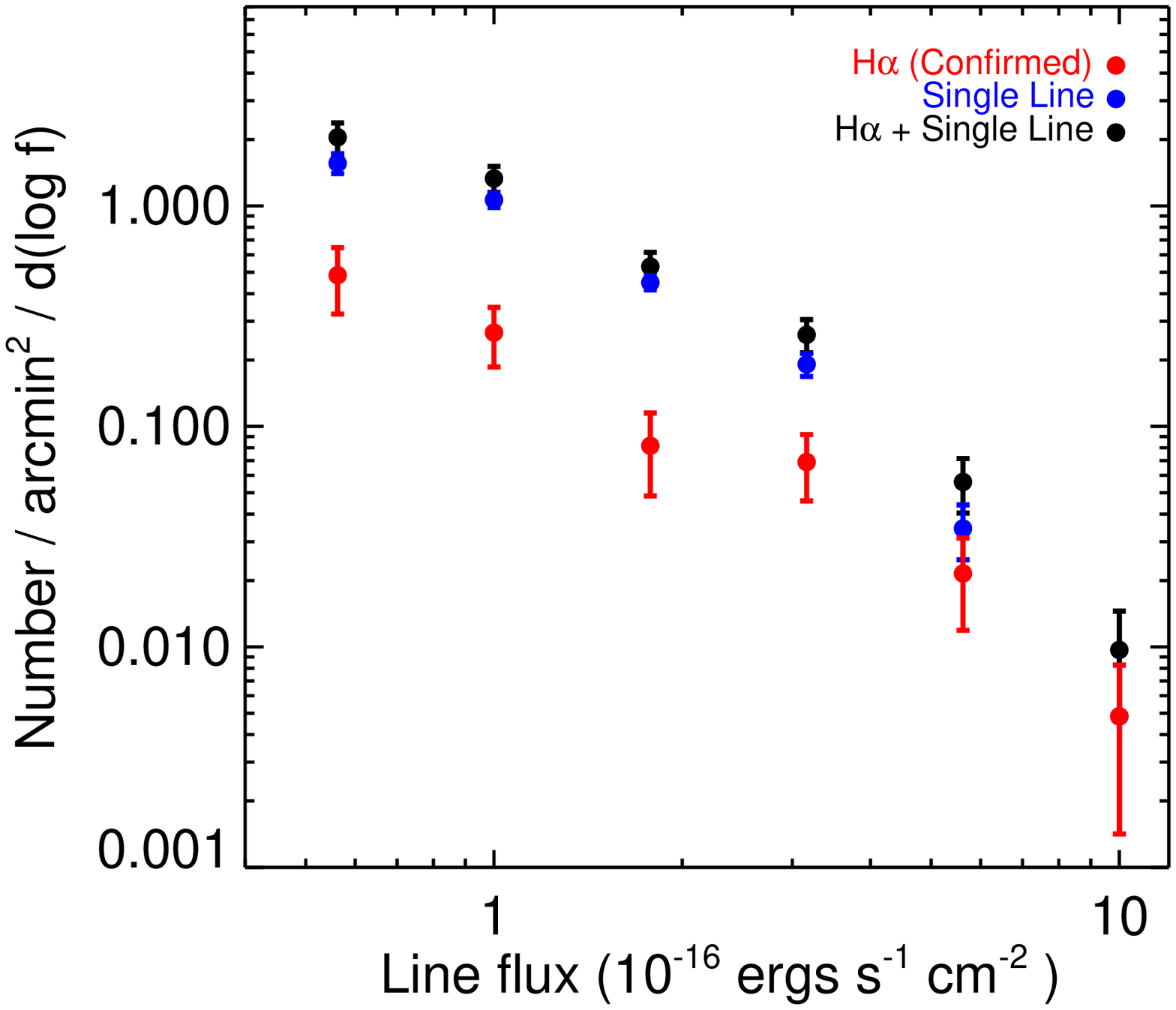}
\hspace{0.2cm}
\includegraphics[width=8.5cm]{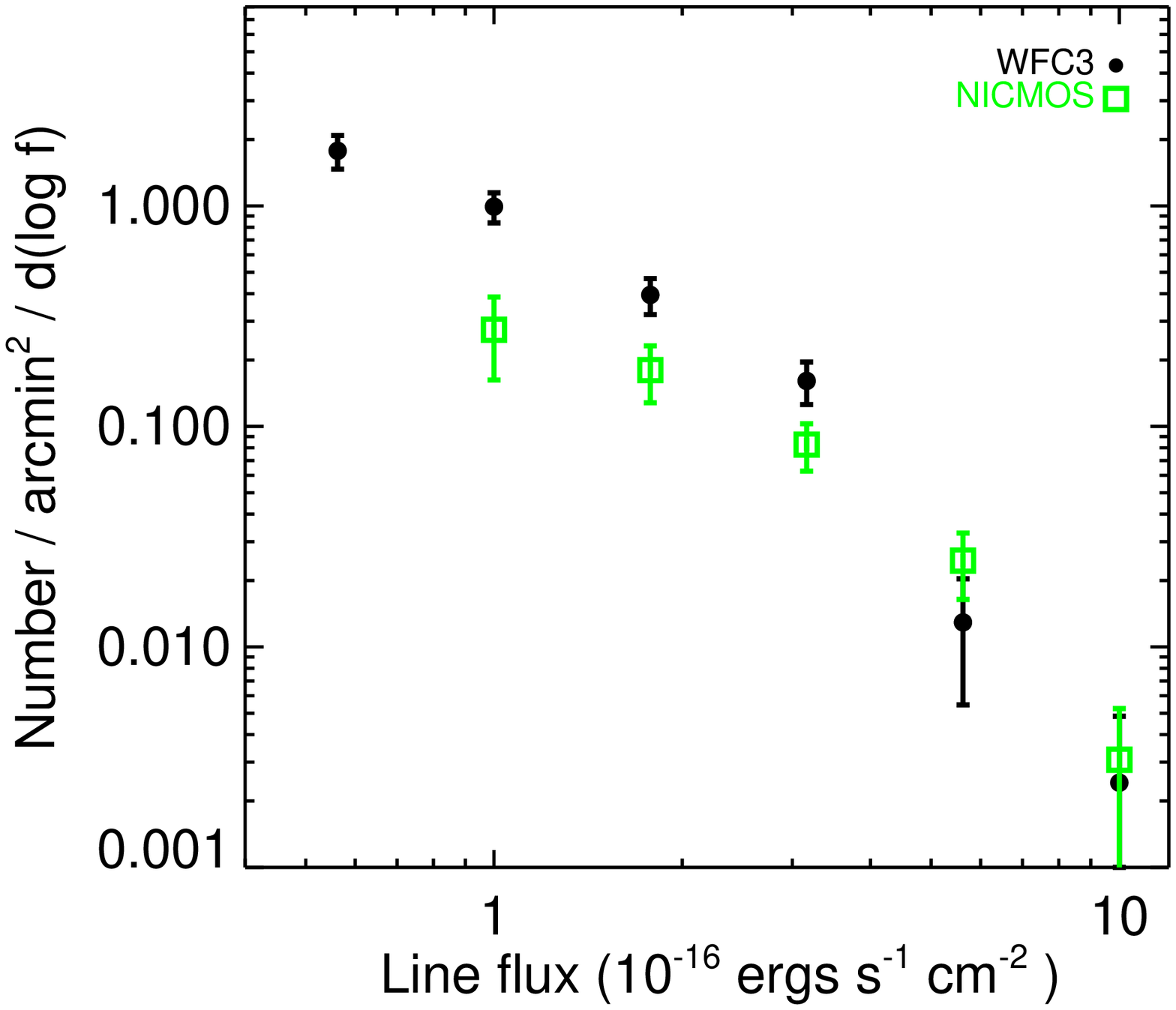}
\caption{{\it Left --}Number counts of \ha\ emitters in the WISP survey. Confirmed \ha\ emitters are shown in red points, single-line emitters in blue and the total counts in black. {\it Right--} 
Number counts of \ha\ and single line emitters in the WISP survey (black points) compared to the NICMOS Parallel grism survey (green squares), tabulated in the overlapping redshift range of  $0.7 < z < 1.5$. Counts are given per $d(logf) = 0.25$. }
\label{fig:ncounts}
\end{figure*}

We next compare these first results to our previous NICMOS \ha\ survey \citep{pmc99,shim09}. With WFC3, we see \ha\ emission
from $0.3 < z < 1.5$, whereas with the NICMOS G141 survey we were sensitive to \ha\ from $0.7 < z < 1.9$.   In the left panel 
of Figure \ref{fig:ncounts}, we plot raw number counts for the \ha\ and single line emitters separately (red and blue points respectively) then for the total (\ha\ + single-line) sample (black). We then compare total counts from both surveys, in the overlapping redshift range of  $0.7 < z< 1.5$ (right panel). At the brighter end, we find a good 
agreement between the WISP survey and the NICMOS grism results, even without including the many
lines which we have identified with a lower signal-to-noise ratio. But WFC3 starts to detect more galaxies than NICMOS at $f \sim 3 \times 10^{-16}$ ergs s$^{-1}$ cm$^{-2}$. Towards the faint-end and the completeness limit of NICMOS,  the surface density of \ha\ emitters detected with WFC3 is a factor of 3 higher. This is because our survey is more complete for extended line-emission and our spectra are sensitive to a very wide range of emission line equivalent widths, from cases where the continuum is hardly detected (EW$_{obs}> 300$\AA, see Figure \ref{fig:examples4}), to low contrast features with EW as low as 15 \AA\ for lines that satisfy S/N = 5$\sigma$.

As Figure \ref{fig:mag_z} illustrates, we routinely detect emission-lines in
galaxies with luminosities well below L$^*$. Many of these very faint
galaxies have strong [OIII]5007 emission and relatively weak [OII] and
\hb\ emission, suggestive of low metal abundances. The near-IR
photometry provided by our F110W and F140W images, and ground-based
follow-up (where practical), will allow us to extend pioneering work
on the mass-metallicity relation at $0.5 < z < 1$\ \citep{savaglio05}\
and $1 < z < 3$\ \citep{erb06} to significantly lower stellar
masses. The mass-metalicity relation is a critical diagnostic of
galaxy formation and evolution models as it is sensitive to a number
of physical process (e.g. infall, SNe feedback, IMF). 

\begin{figure*}[htbp]
\includegraphics[width=9.2cm]{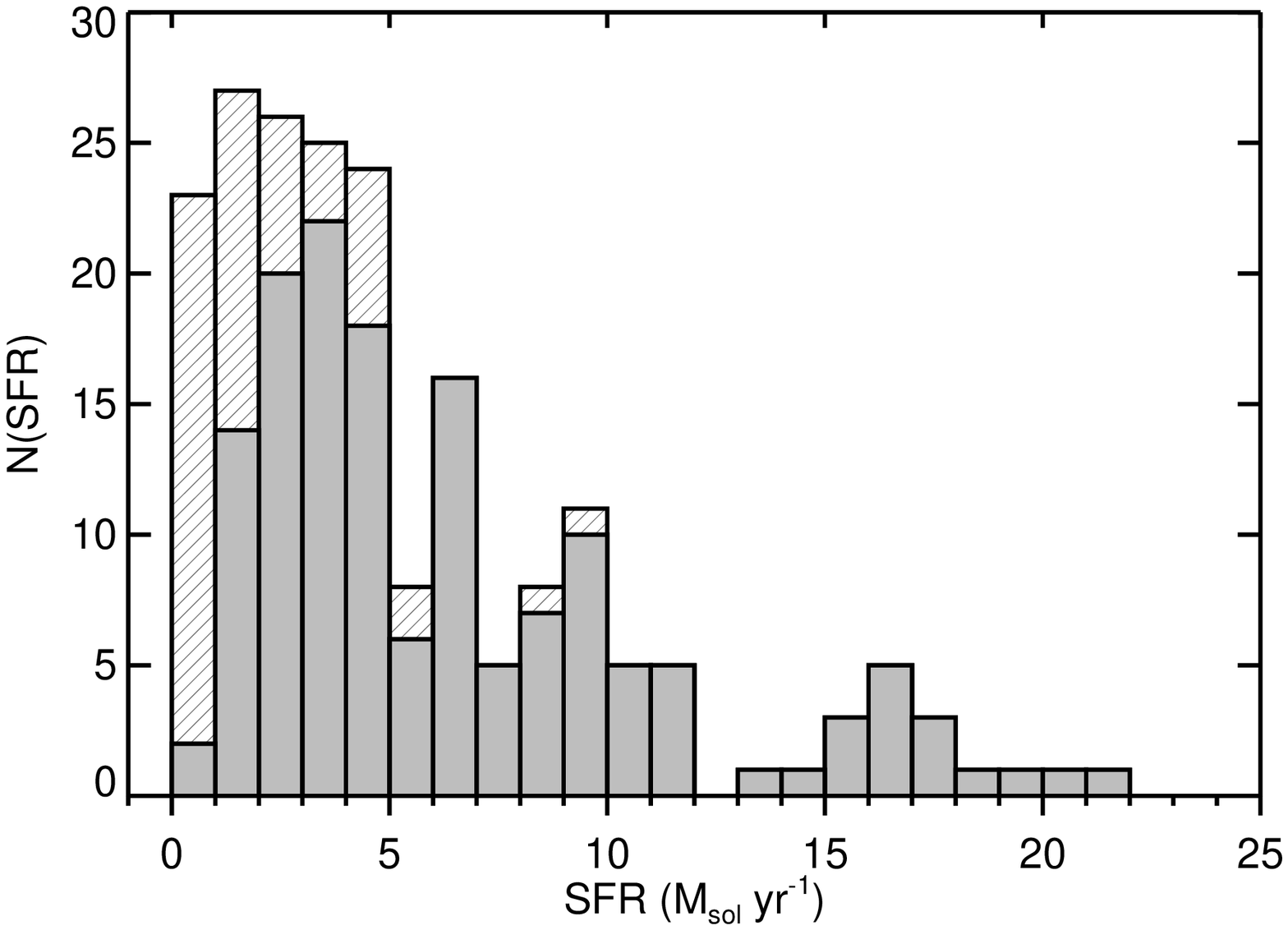}
\includegraphics[width=9.2cm]{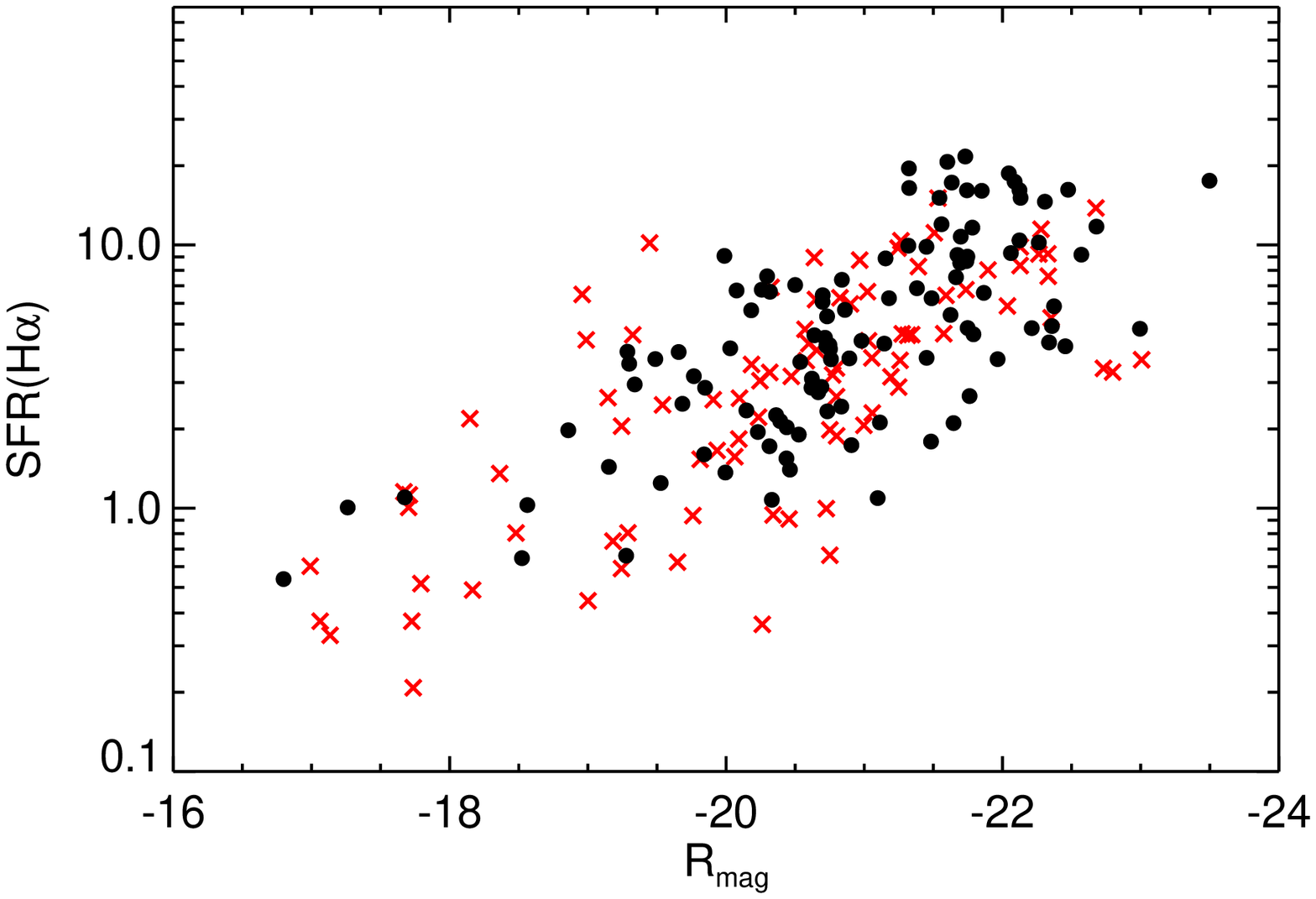}
\caption{ {\it Left --} Observed star formation rate distribution of our \ha\ emitters. The \ha\ flux is reduced to account for 20\% of [\ion{N}{2}] contamination and converted to SFR using \citet{kennicutt98} relation. The sample is split into high-z (shaded area) and low-z (hatched) sub-samples, where the separation between the two categories is  z=0.675.  {\it Right --}  SFR as a function of absolute R-band magnitude derived from F110W and F140W photometry for \ha\ emission below and above 1.25 microns respectively. Black points represent confirmed \ha\ emitters and red crosses the single-line galaxies.}
\label{fig:sfr}
\end{figure*}

In Figure \ref{fig:sfr} we plot on the left panel the observed star formation rate distribution of our \ha\ sample derived from the \ha\ luminosity using the \cite{kennicutt98} conversion. Following \citet{shapley05}, we corrected the \ha\ fluxes for 20\% average [NII]6584+6548 contribution, but no extinction correction is applied. On the right panel we plot SFR(\ha) versus rest-frame R-band luminosity. For rest wavelength R magnitudes
we used the F110W photometry if the observed wavelength of \ha\ is
$<$ 1.25 \mic\ ($z<0.9$) and the F140W photometry for \ha\ $>$ 1.25
\mic\ ($z>0.9$), applying only the ($1+z$) correction (without any k-correction). There is no
extinction correction for the R-band magnitudes.  The median SFR is 4
M$_{\odot}$/year, and the median M$_R$ of line-detected galaxies is
-20.75, one and half magnitudes below L$^{\star}$ for z $\sim 1.0$ \ \citep{chen03,colbert06}.
At the lower redshifts, especially
for compact objects, our survey is able to detect lines in
extraordinarily faint galaxies (M$_{R} > -19$), and star formation rates less than 1 solar-masse per year.



\begin{figure*}
\centering
\includegraphics[width=12cm]{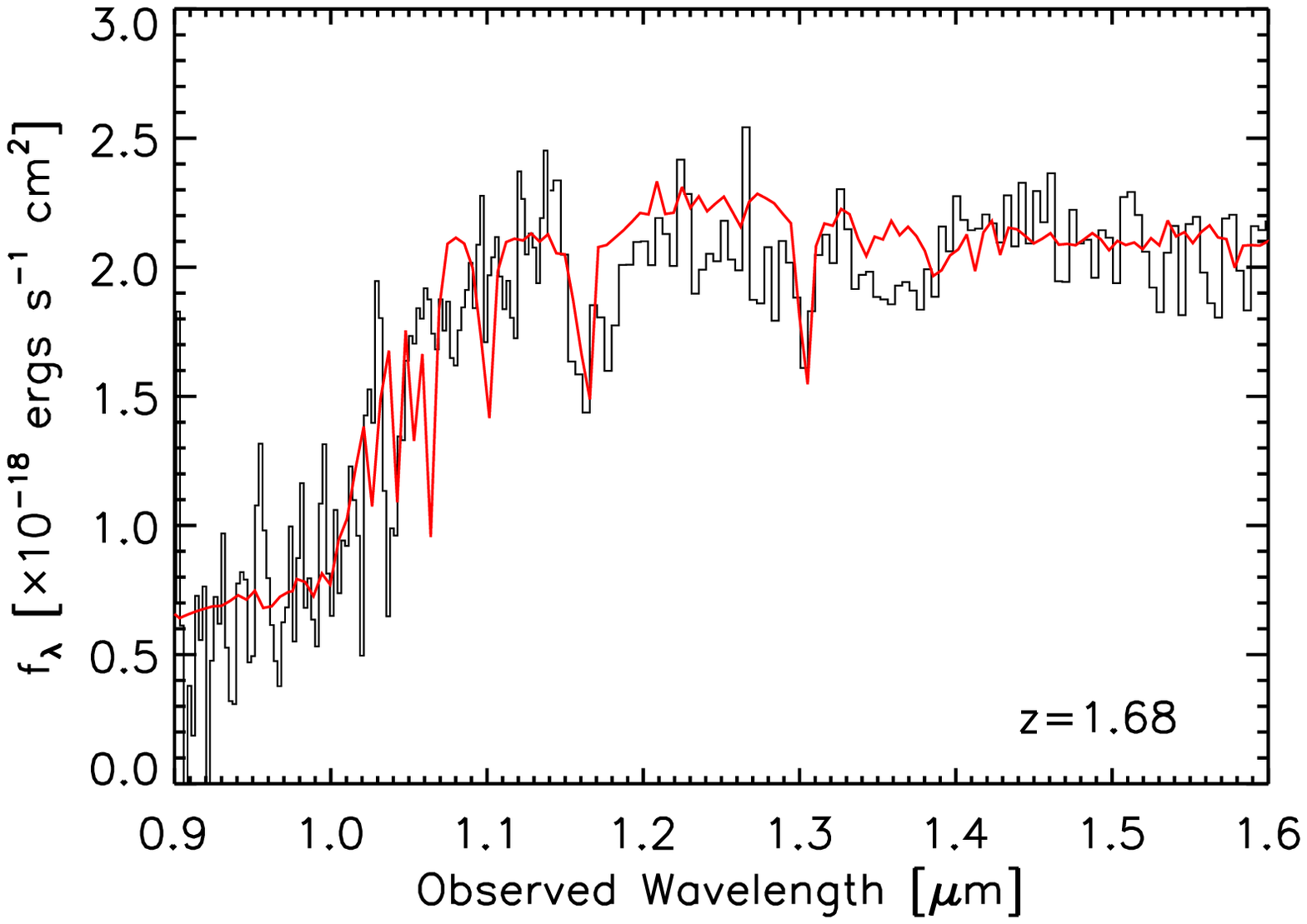}\\
\caption{Example {\sc V}: Spectrum of a post-starburst galaxy at $z=1.68$.  A clear Balmer break can be seen in the G102 spectrum between $1.0<\lambda<1.1 \mu$m as well as Balmer absorption lines in the G141 spectrum at 1.17 and 1.3 $\mu$m.  Plotted (in red) is a Bruzual \& Charlot (2003) stellar population model with an age of 600 Myr (e-folding time of 100 Myr), $A_{V} = 1.2$ mags (assuming a Calzetti reddening curve), and a Salpeter IMF. }
\label{fig:examples5}
\end{figure*}

\subsection {Absorption Line Spectra}

Although we have so far limited the analysis to emission lines, the
broad wavelength coverage and high sensitivity of the WISP
survey leads also to a wealth of information from continuum
spectra. The 4000\AA\ break and the convergence of the Balmer series at
3650\AA, provide two of the best low-resolution diagnostics for
redshift and stellar populations.  Continuum spectral diagnostics,
particularly at $z > 1$, provide constraints on the mean age of the
stellar population and the time since the last major episode of star
formation.  Spectroscopy of red galaxies from the ground in the
visible is limited to objects with $z < 1.8$ and $I < 25$\
\citep[e.g.][]{pmc04,cimatti04,doherty05}.
Near-IR spectroscopy can probe to higher redshifts, but
this is quite challenging \citep[e.g.][]{kriek08} and has
been limited to small samples to date \citep[e.g.][]{doherty06}. The WFC3 grisms are providing
low-resolution continuum spectroscopy of red galaxies in large numbers
with great sensitivity \citep[e.g.][]{vandokkum10}. In Figure~\ref{fig:examples5} we
show an example of a post-starburst galaxy at redshift
$z=1.68$. A clear Balmer break is seen in the G102 spectrum, as well as absorption lines in the G141 part. A stellar population model is overplotted, giving an age of 600 Myr and an extinction of A$_{V} = 1.2$. This object has been pre-identified searching for
galaxies with $H\textunderscore140_{AB}<22$ and $J\textunderscore110-H\textunderscore140>0.6$. This selection isolates
bright passive galaxies at $z>1$. 


\section{Discussion and Summary}\label{sec:discuss}

We have presented the first analysis of the WISP Survey of slitless
spectroscopy with the NIR channel of WFC3, taken in pure parallel
mode.  In the first 19 fields, we surveyed 63 arcmin$^{2}$.
At the highest confidence level we have 
detected 229 emission line objects.  We reach typical $5\sigma$\
sensitivities of $f\sim5\times 10^{-17}$~ergs s$^{-1}$~cm$^{-2}$\ for
compact lines. 

While wide-field narrow-band imaging surveys from the ground \citep[e.g.][]{sorbal09,momcheva10}
yield samples of H$\alpha$ emitters that are larger than ours and near-IR
slit spectroscopy of selected targets can yield emission-line ratios to faint levels
(e.g. Pettini et al. 2001), the WFC3 grisms offer a unique opportunity to obtain continuous
spectral coverage in the $0.8 - 1.7$\mic\ region with high sensitivity. Our
early results show that we can identify star forming galaxies with space densities
comparable to, or larger than, those identified with color selection techniques
(e.g. BM/BX or LBG selection; Adelberger et al. 2004) to similar star formation rate thresholds (Shapley et al. 2005, Erb et al. 2006). The typical emission-line
galaxy at $z > 1.3$ in our sample has strong [OIII]5007,4959 emission and a large
[\ion{O}{3}]/H$\beta$ ratio, similar to the ratios seen in the small number of LBGs at $z>1.5$ with rest-frame optical spectra (Pettini et al. 2001, Teplitz et al. 2000, Lemoine-Busserolle et al. 2003, Maiolino et al. 2008, Hainline et al. 2009). Coupled with the high [\ion{O}{3}]/H$\alpha$ ratio
and weak [\ion{O}{2}]3727 these suggest low metal abundances. Future papers from our
survey will examine line ratios and their implications for abundances and
ionization sources. With the support of our ongoing ground-based, follow-up observing programs, we will be able to measure the mass-metallicity relation at crucial 
intermediate redshifts, to derive the \ha\ luminosity function at different redshifts and to study the evolution of the star formation rate density.

Many of our fields contain emission-line objects with extremely high equivalent
widths (cf. Figure \ref{fig:examples4}). These strongly star forming emission-line dominated objects
are reminiscent of the ultra-strong emission-line (USEL) galaxies at
$0<z<1$\ discovered by \cite{kakazu07} \citep[see also][]{hu09} and the ``green pea'' galaxies found
in SDSS by Cardamone et al. (2009). These objects 
have metallicities comparable to the lowest observed in local sources and
may be dwarf galaxies undergoing their first major star formation episode.

The sensitivity and resolution of the G102 grism also provide the
opportunity to search for Ly$\alpha$\ emission lines in addition to
the rest-frame optical emission lines discussed here. We have excluded
such analysis from the present paper given the limitations of the
initial, visual selection of faint emission lines. After further
analysis, WFC3 will offer a significant new approach to high-redshift
($z> 6$) searches for Ly$\alpha$. Though extensive ground-based
narrowband searches for Ly$\alpha$-emitters have been performed up to
$z=7$, with the most distant confirmed source at $z=6.96$ \citep[][]{iye06}, it is difficult beyond this redshift to find wavelengths with
both high sky transmission and low sky emission. Thus, many surveys
are limited to a very small range at $z=7.7$ or $z=8.8$\
\citep[e.g.][]{sorbal09}, without any reliable Ly$\alpha$ emission-line candidates.

In addition to the WISP survey, two large observing programs with the WFC3 grisms will
contribute significant results. First, a G141 survey of the GOODS
North fields is underway (GO-11600; PI=Weiner), reaching depths
comparable or slightly better than WISP.  Secondly, the CANDELS
multi-cycle treasury program (GO-12060 through 12064; PI=Faber and Ferguson) will
obtain deep spectroscopy of selected pointings targetted to follow-up
supernovae found in direct imaging. The WISP survey has some
limitations compared to these programs due to the parallel mode,
including the lack of dithering, but
our first analysis demonstrates the great
efficiency of this use of the telescope.  
WFC3 parallels can obtain large spectroscopic
samples of emission-line objects during the peak epoch of star
formation, free from the most common selection biases.



\acknowledgments
We gratefully acknowledge the dedicated efforts of several staff members at the Space
Telescope Institute to make the new Grism Parallel observations as successful as
possible.  In particular, we thank Galina Soutchkova, Iain Neill Reid, Larry Petro, Karla Peterson,
Denise Taylor, Kenneth Sembach,
Bill Workman, Claus Leitherer, John MacKenty, Howard Bushouse, Ron Downes and Alan Patterson. We thank the ST-ECF team of Martin K\"ummel, Harald Kuntschner and Jeremy Walsh for their help with the data reduction and spectrum extraction processes and advice about the WFC3 calibration status. We also acknowledge the important contributions to supporting parallel observations by the late Roger Doxey.

\pagebreak


\end{document}